\documentclass[a4paper,fleqn,usenatbib]{mnras}
\usepackage{newtxtext,newtxmath}
\usepackage[T1]{fontenc}
\usepackage{ae,aecompl}

\usepackage{graphicx}
\usepackage{amsmath,amssymb}
\usepackage{gensymb}
\usepackage{hyperref}

\newcommand{\lya}{Ly$\alpha$}

\newcommand{\civ}{\hbox{C\,{\sc iv}}}
\newcommand{\oiii}{\hbox{O\,{\sc iii}}}

\newcommand{\nev}{\hbox{Ne\,{\sc v}}}

\newcommand{\hii}{\hbox{H\,{\sc ii}}}

\newcommand{\heii}{\hbox{He\,{\sc ii}}}

\newcommand{\heplus}{\ensuremath{\rm He^+}}

\newcommand{\mbh}{\ensuremath{\rm M_{\mathrm{BH}}}}
\newcommand{\rmax}{\ensuremath{\rm R_{\mathrm{max}}}}
\newcommand{\lxb}{\ensuremath{\rm L_{\mathrm{X, 0.5-8\, keV}}}}

\providecommand{\e}[1]{\ensuremath{\times 10^{#1}}}
\newcommand{\unit}[1]{\ensuremath{\, \mathrm{#1}}}

\DeclareRobustCommand{\ion}[2]{\relax\ifmmode\ifx\testbx\f@series{\mathbf{#1\,\mathsc{#2}}}\else{\mathrm{#1\,\mathsc{#2}}}\fi\else\textup{#1\,{\mdseries\textsc{#2}}}\fi}

\newcommand{\hstcos}{\textit{HST}/COS}

\newcommand{\chandra}{\textit{Chandra}}

\newcommand{\cloudy}{\textsc{cloudy}}

\defcitealias{senchynaUltravioletSpectraExtreme2017}{S17}
\defcitealias{senchynaExtremelyMetalpoorGalaxies2019}{S19}
\defcitealias{brorbyEnhancedXrayEmission2016}{B16}

\title[HMXBs and nebular \heii{}]{High-mass X-ray binaries in nearby metal-poor galaxies: on the contribution to nebular \heii{} emission}
\author[P. Senchyna et al.]{
    Peter Senchyna$^{1}$\footnotemark[1],
    Daniel P. Stark$^{1}$,
    Jordan Mirocha$^{2,3}$, 
    Amy E.\ Reines$^{4}$, \newauthor
    St\'{e}phane Charlot$^{5}$,
    Tucker Jones$^{6,7}$,
    and John S.\ Mulchaey$^{8}$
    \\
    \vspace{0.1in}\\
    $^{1}$ Steward Observatory, University of Arizona, 933 N Cherry Ave, Tucson, AZ 85721 USA \\
    $^{2}$ Department of Physics and McGill Space Institute, McGill University, 3550 rue University, Montreal, QC, H3A 2T8, Canada \\
    $^{3}$ CITA National Fellow \\
    $^{4}$ eXtreme Gravity Institute, Department of Physics, Montana State University, Bozeman, MT 59717, USA \\
    $^{5}$ Sorbonne Universit\'{e}, CNRS, UMR7095, Institut d'Astrophysique de Paris, F-75014, Paris, France \\
    $^{6}$ Department of Physics, University of California Davis, 1 Shields Avenue, Davis, CA 95616, USA \\
    $^{7}$ Institute of Astronomy, University of Hawaii, 2680 Woodlawn Drive, Honolulu, HI 96822, USA \\
    $^{8}$ Observatories of the Carnegie Institution of Washington, 813 Santa Barbara Street, Pasadena, CA 91101, USA \\
}

\date{Accepted XXX. Received YYY; in original form ZZZ}

\pubyear{2020}

\begin{document}
\label{firstpage}
\pagerange{\pageref{firstpage}--\pageref{lastpage}}
\maketitle

\begin{abstract}
    Despite significant progress both observationally and theoretically, the origin of high-ionization nebular \heii{} emission in galaxies dominated by stellar photoionization remains unclear.
    Accretion-powered radiation from high-mass X-ray binaries (HMXBs) is still one of the leading proposed explanations for the missing $\mathrm{He^+}$-ionizing photons, but this scenario has yet to be conclusively tested.
    In this paper, we present nebular line predictions from a grid of photoionization models with input SEDs containing the joint contribution of both stellar atmospheres and a multi-color disk model for HMXBs.
    This grid demonstrates that HMXBs are inefficient producers of the photons necessary to power \heii{}, and can only boost this line substantially in galaxies with HMXB populations large enough to power X-ray luminosities of $10^{42}$ erg/s per unit star formation rate (SFR).
    To test this, we assemble a sample of eleven low-redshift star-forming galaxies with high-quality constraints on both X-ray emission from \chandra{} and \heii{} emission from deep optical spectra, including new observations with the MMT.
    These data reveal that the HMXB populations of these nearby systems are insufficient to account for the observed \heii{} strengths, with typical X-ray luminosities or upper limits thereon of only $10^{40}$--$10^{41}$ erg/s per SFR.
    This indicates that HMXBs are not the dominant source of $\mathrm{He^+}$ ionization in these metal-poor star-forming galaxies.
    We suggest that the solution may instead reside in revisions to stellar wind predictions, softer X-ray sources, or very hot products of binary evolution at low metallicity.
\end{abstract}


\footnotetext[1]{E-mail: senchp@email.arizona.edu}

\section{Introduction}

As the nearest collections of young and very metal-poor stars in the universe, local star-forming dwarf galaxies represent a crucial testbed for models of stellar populations at low metallicity.
Nebular emission lines from highly-ionized gas provide a window onto the shape of the ionizing spectra of these systems in the extreme ultraviolet (EUV, 10--100 eV).
The EUV potentially contains contributions from both the hottest stars and emission from gas heated by compact object accretion and shocks, all of which are highly uncertain at very low-metallicity.
The promise of obtaining quantitative constraints on these processes with deep spectroscopy is alluring, but physically interpreting this nebular emission has proved challenging.

In particular, the presence of nebular \heii{} emission in nearby star-forming regions remains a puzzle three decades after it was first noted \citep[][and references therein]{garnettHeIIEmission1991}.
The difficulty in understanding this emission is rooted in the very high energy of the photons necessary to doubly-ionize helium ($>54.4$ eV, or $>4$ Ryd) and thus power the recombination spectrum of \heii{}; most notably, \heii{}$\lambda 4686$ \AA{} and $\lambda 1640$ \AA{}.
Theoretical stellar atmosphere models generally predict very little emergent flux beyond 54.4 eV, as $\mathrm{He^+}$ ionization in the atmosphere and expanding winds of massive stars introduces a strong absorption edge at this energy \citep[e.g.][]{gablerUnitifiedNLTEModel1989,pauldrachRadiationdrivenWindsHot2001,lanzGridNonLTELineblanketed2003,pulsAtmosphericNLTEmodelsSpectroscopic2005}.
In order to account for this apparent excess in flux at the $\mathrm{He}^+$-ionizing edge relative to the stellar models, \citet{garnettHeIIEmission1991} proposed two alternative non-stellar sources of ionizing radiation: radiative shocks and X-ray binaries.

Massive progress has since been made in expanding the sample of nebular \heii{} detections both in the local Universe \citep[e.g.][]{thuanHighIonizationEmissionMetaldeficient2005,kehrigGeminiGMOSSpectroscopy2011,shiraziStronglyStarForming2012,kehrigExtendedHeII2015,senchynaUltravioletSpectraExtreme2017,kehrigExtendedHeII2018,bergIntenseIVHe2019} and at $z\sim 1$--4 \citep[e.g.][]{erbPhysicalConditionsYoung2010,cassataHeIIEmitters2013a,bergWindowEarliestStar2018,nanayakkaraExploringHeII2019}.
This body of observational evidence has made clear that nebular \heii{} is strongly metallicity-dependent, and is likely ubiquitous among star-forming systems at metallicities $12+\log\mathrm{O/H}<7.7$ \citep[equivalently, $Z/Z_\odot<0.1$; e.g.][]{senchynaPhotometricIdentificationMMT2019a,senchynaExtremelyMetalpoorGalaxies2019}.
Yet despite commensurate advances in stellar modeling, including state-of-the-art treatments of atmospheres, binarity, and rotation \citep[e.g.][]{,szecsiLowmetallicityMassiveSingle2015,gotbergIonizingSpectraStars2017,eldridgeBinaryPopulationSpectral2017,gotbergSpectralModelsBinary2018,stanwayInitialMassFunction2019,kubatovaLowmetallicityMassiveSingle2019}, there is as-yet no clear solution to the apparent modeling deficit of hard ionizing photons.

In recent years, the \chandra{} X-ray Observatory has revolutionized our understanding of X-ray binaries, enabling a reappraisal of their possible role in providing the missing ionizing photons.
High-mass X-ray binaries (HMXBs), which power hard X-ray emission via accretion from a massive stellar companion onto a compact object (black hole or neutron star), have been shown to power nebular \heii{} in cases where they act as the sole ionizing source in a nebula \citep{pakullDetectionXrayionizedNebula1986a,kaaretHighresolutionImagingHe2004,gutierrezOpticalStudyHyperluminous2014}.
While it has long been established that HMXBs dominate the hard X-ray flux ($\gtrsim 1$ keV) of actively star-forming galaxies in the local Universe \citep[e.g.][]{grimmHighmassXrayBinaries2003,mineoXrayEmissionStarforming2012}, early studies largely ignored very low-metallicity galaxies due to their faintness.
Now, strong evidence has arisen that the X-ray luminosity per unit of star formation rate increases by nearly an order of magnitude in extremely metal-poor galaxies ($Z/Z_\odot \lesssim 0.1$) relative to those at near-solar metallicity \citep{mapelliUltraluminousXraySources2010a,prestwichUltraluminousXRaySources2013,basu-zychEvidenceElevatedXRay2013,brorbyXrayBinaryFormation2014,dounaMetallicityDependenceHighmass2015,brorbyEnhancedXrayEmission2016,brorbyXraysGreenPea2017}.
Theoretical models reproduce this general trend, suggesting that it is likely driven by the weaker stellar winds driven at low metallicity \citep[e.g.][]{drayMetallicityDependenceHighmass2006,lindenEffectStarburstMetallicity2010,fragosXRAYBINARYEVOLUTION2013}.
In particular, weaker winds lead to both more massive black holes and a higher incidence of HMXB systems undergoing Roche lobe overflow accretion, which both tend to produce more luminous X-ray binary populations.

The similarity of the metallicity dependence of HMXBs to that of nebular \heii{} motivated \citet{schaererXrayBinariesOrigin2019} to revisit the possibility that HMXBs may be solely responsible for this emission line in metal-poor star-forming galaxies.
By assuming that the nebular \heii{} in IZw18 NW is entirely powered by the X-ray source observed in that cluster \citep[c.f.][]{lebouteillerNeutralGasHeating2017}, \citet{schaererXrayBinariesOrigin2019} derive a scaling relationship between \heii{} flux and hard X-ray luminosity.
Applying this to the population synthesis models of \citet{fragosXRAYBINARYEVOLUTION2013,fragosEnergyFeedbackXRay2013a}, the authors demonstrate that the metallicity-dependent HMXB population produces a trend in the strength of \heii{} relative to H$\beta$ similar to that observed.
Though suggestive, the model presented by \citet{schaererXrayBinariesOrigin2019} has yet to clear two critical hurdles.
First, it has yet to be demonstrated through full photoionization modeling that an HMXB spectrum can successfully power nebular \heii{} when combined with stellar ionizing flux.
Second, a detailed galaxy-by-galaxy investigation of both \heii{} and X-ray constraints has not yet been conducted for more than a handful of systems \citep[e.g.][]{thuanHighIonizationEmissionMetaldeficient2005,kehrigExtendedHeII2018}.

In this paper, we will test the hypothesis that HMXBs dominate the production of \heii{} in star-forming galaxies from both a theoretical and observational perspective.
First, we produce a grid of photoionization models with input SEDs reflecting the joint impact of young stellar populations and a variable HMXB contribution, and examine the impact of the latter on the predicted nebular line spectrum (Section~\ref{sec:model}).
Then, leveraging recent work targeting metal-poor galaxies with both \chandra{} and high-resolution optical spectroscopy including data from Keck and new measurements from the MMT 6.5m telescope, we assemble a sample of eleven star-forming regions with robust constraints on both X-ray emission and nebular \heii{} in Section~\ref{sec:obs}.
In Section~\ref{sec:datajointres} we examine these observational constraints in the context of the photoionization model results, providing a stringent empirical test of the claim that HXMBs dominate production of $\mathrm{He^+}$-ionizing photons in these systems.
We conclude in Section~\ref{sec:summary}.

\section{Modeling Gas Photoionized by Stars and HMXBs}
\label{sec:model}

In order to investigate the possibility that \heii{} or other high-ionization emission lines are powered by high-mass X-ray binaries, we construct a semi-empirical framework to model this scenario explicitly.
This requires us first to construct a model SED representing the combined ionizing spectrum of both massive stars and the emission from HMXB accretion disks.
Then, we use a photoionization modeling framework to simulate the reprocessing of this ionizing spectrum through surrounding gas, producing predictions about the resultant nebular emission.
We describe these two steps below in Section~\ref{subsec:modelmeth}, then discuss the results and predictions of this modeling in Section~\ref{subsec:modelres}.

\subsection{Methodology}
\label{subsec:modelmeth}

Our primary goal in this section is to assess the magnitude of the effect on nebular emission lines introduced by adding a HMXB spectrum to a stellar population model.
While there are many existing prescriptions for full galaxy SED modeling in the literature \citep[see e.g.][]{charlotNebularEmissionStarforming2001,gutkinModellingNebularEmission2016,chevallardModellingInterpretingSpectral2016,lejaDerivingPhysicalProperties2017,bylerNebularContinuumLine2017,fiocPEGASECodeModeling2019}, none yet account explicitly for the impact of HMXBs on nebular gas emission.
We describe out methodology in this subsection with comparison to other approaches in the literature where relevant.

\subsubsection{Construction of the Spectral Energy Distribution}

First, we describe how we construct the joint SED representing the ionizing flux emitted by massive stars and HMXBs.
Since strong interstellar absorption precludes the direct observation of the EUV spectrum of massive stars or HMXBs, we must rely on models calibrated at higher or lower energies to predict the SED in this energy regime.

There are a variety of stellar population synthesis frameworks in the literature designed to predict the emergent flux from stars themselves \citep[for a review, see][]{conroyModelingPanchromaticSpectral2013}.
As we will discuss further later in the paper, the shape of model stellar spectra in the EUV beyond the Lyman limit is highly uncertain, as these photons cannot be directly observed for any individual hot OB stars.
In particular, the emergent flux from stars at and just beyond the $\mathrm{He}^+$-ionizing edge (54.4 eV, or 228 \AA{}) is subject to substantial systematic uncertainties from both atmosphere and evolutionary models, and varies significantly between different population synthesis prescriptions.
For instance, mass transfer can strip the donor star of its outer hydrogen layers, and potentially spin-up the acceptor star sufficiently to change its evolution, both of which can enhance the escape rate for $\mathrm{He}^+$-ionizing photons which are easily blocked by the outer layers or dense winds of `typical' massive stars \citep[e.g.][]{eldridgeEffectStellarEvolution2012,szecsiLowmetallicityMassiveSingle2015,gotbergSpectralModelsBinary2018,stanwayInitialMassFunction2019,gotbergImpactStarsStripped2019a}.
Accounting for high ZAMS rotation rates can also boost stellar flux in the EUV \citep[e.g.][]{maederEvolutionRotatingStars2000,vazquezModelsMassiveStellar2007,levesqueEffectsStellarRotation2012,bylerNebularContinuumLine2017}.
Significant uncertainties remain in the specific treatment of all of these factors, and they have not yet been considered simultaneously in a full population synthesis prescription.

For the purposes of this paper, we are primarily interested in constraining the potential impact of HMXBs on nebular emission when added to a stellar population.
Thus, we focus solely on the latest version of the BPASS models accounting for binary evolutionary effects \cite[2.2, described in][]{stanwayReevaluatingOldStellar2018}.
These models incorporate prescriptions for some binary mass transfer processes which can have a substantial impact on emergent hard ionizing flux.
It is important to note that modifications to our treatment of the stellar population that tend to increase the stellar contribution to flux at the $\mathrm{He^+}$-ionizing edge will decrease the relative impact of the HMXB spectrum.
The BPASS flux predictions in the EUV are fairly representative of the state-of-the-art population synthesis codes, and in particular are lower than the newest results from the modified \citet{bruzualStellarPopulationSynthesis2003} models (Charlot \& Bruzual, in-prep.) incorporating newer theoretical stellar atmospheres for massive main-sequence and Wolf-Rayet stars \citep[see][]{stanwayInitialMassFunction2019,platConstraintsProductionEscape2019}.

The ionizing flux of of a purely-stellar SED depends strongly on the assumed IMF, star formation history, and stellar metallicity.
Since we are interested primarily in the HMXB contribution, we will make reasonable assumptions about the IMF and star formation history while leaving metallicity as a free parameter.
We assume the fiducial BPASS IMF, which consists of a broken power-law with a \citet{salpeterLuminosityFunctionStellar1955} slope ($-2.35$) over the mass range 0.5--300 $M_\odot$.
Extreme IMFs can enhance the $\mathrm{He^+}$-ionizing flux achieved, though not sufficiently to alone explain the strongest \heii{} emission observed \citep{stanwayInitialMassFunction2019}.
We assume a constant star formation history that has proceeded for sufficiently long for the SED to stabilize ($100$ Myr), noting that instead adopting younger ages or adding young bursts will act to enhance the relative flux beyond 54.4 eV per unit SFR by increasing the number of early O and Wolf-Rayet stars \citep[e.g.][]{shiraziStronglyStarForming2012,chisholmConstrainingMetallicitiesAges2019a}.
Adopting a more extreme IMF or a star formation history weighted to younger ages would both increase the amount of hard ionizing flux from the stellar population at fixed SFR and reduce the relative impact of the HMXB spectrum on nebular line emission.
Our relatively conservative assumptions about the stellar population synthesis prescription, IMF, and star formation history are chosen to provide a reasonable first estimate of the effect on nebular lines of adding HMXBs to a stellar ionizing spectrum.

With the stellar model prescription, IMF, and star formation history fixed, the most important variable affecting the ionizing flux of the stars themselves is then the stellar metallicity.
As the bulk metallicity of a stellar population is lowered, reduced opacities in the stellar interior and atmosphere lead to both hotter temperature evolution for massive main-sequence stars and dramatically weakened stellar winds.
These factors both directly lead to harder ionizing spectra for individual metal-poor stars, and the reduced impact of stellar wind mass loss at low metallicity can result in rotational and binary evolutionary effects playing a more prominent role in producing very hot stars.
Models of these evolutionary stages are still highly uncertain and are not uniformly included in population synthesis predictions \citep[e.g.][and references therein]{szecsiLowmetallicityMassiveSingle2015,stanwayStellarPopulationEffects2016,gotbergImpactStarsStripped2019a}.
We allow the stellar metallicity to vary from $Z=0.020$ (solar, $Z_\odot$) to $Z=0.001$ ($Z/Z_\odot = 5\%$), the lowest metallicity provided by the BPASSv2.2 grids.
We note that adopting lower stellar metallicities would harden the stellar spectrum and further increase the maximal nebular \heii{} flux powered by the stars alone.
However, this metallicity range encompasses the full range of gas-phase oxygen abundances in our observational sample (assuming solar $\alpha$/Fe; Table~\ref{tab:obssumm}); and as already discussed, such lower metallicities would further diminish the relative impact of HMXBs to \heii{}.

\begin{figure}
    \includegraphics[width=0.5\textwidth]{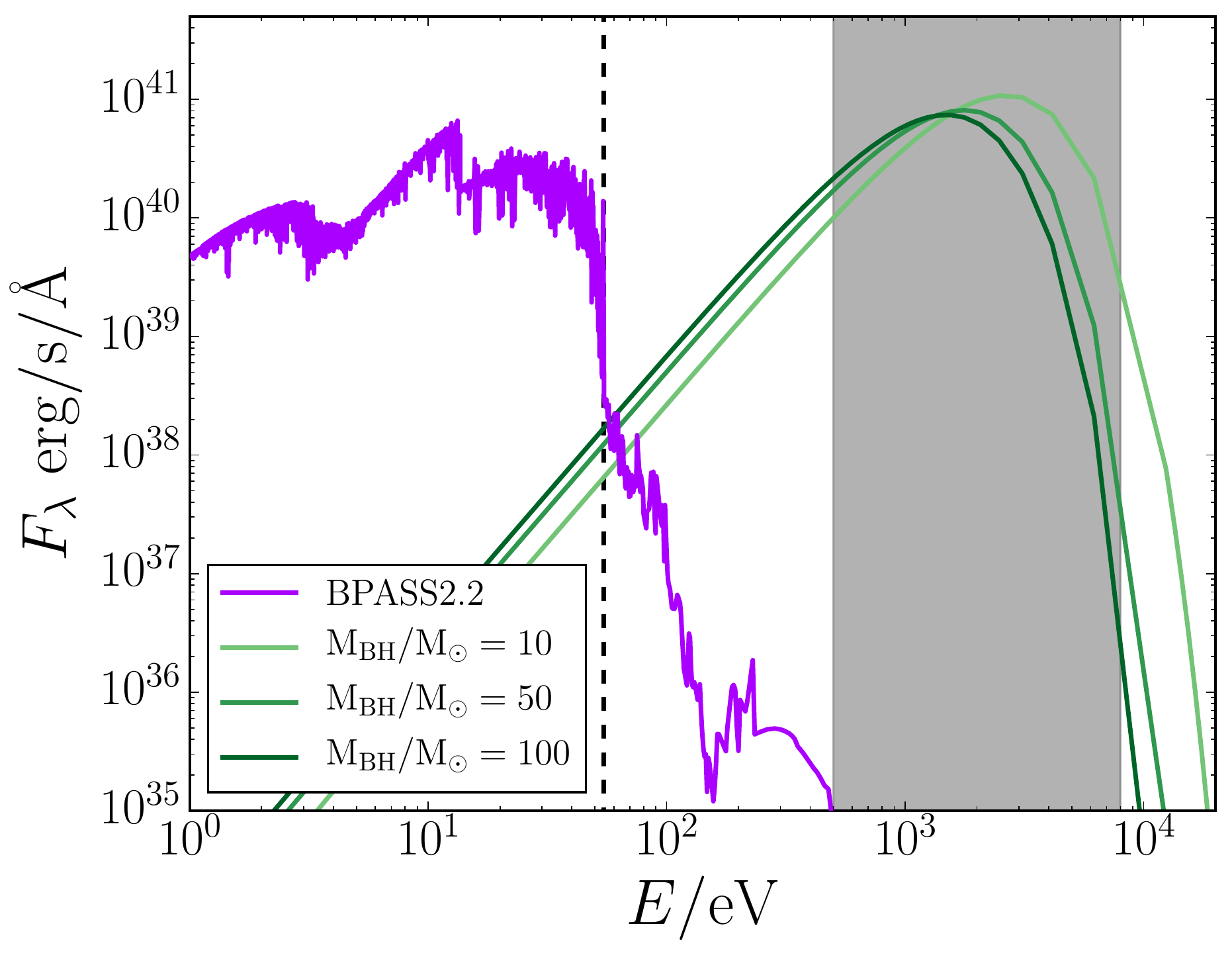}
    \caption{
        Representative SEDs for a model stellar population (BPASS v2.2, constant star formation at $1 \mathrm{M_\odot/yr}$, $Z=0.004$) and a model HMXB (multi-color disk spectrum, $\rmax{}=10^4$ cm, with $\mbh{}/\mathrm{M_\odot}=10$--100).
        The HXMB spectrum is normalized in the 0.5--8 keV range (shaded grey) to the median X-ray production efficiency adopted in our grid: $\lxb{}/\mathrm{SFR}=10^{42} \mathrm{erg/s/(M_\odot/yr)}$.
        At this X-ray normalization, the Rayleigh-Jeans tail of the HMXB spectrum is within an order of magnitude of the stellar SED at the \heplus{}-ionizing edge (black dashed line).
    }
    \label{fig:sedcomp}
\end{figure}

In contrast to the stellar ionizing spectrum which drops-off at energies above 54.4 eV due to absorption in stellar winds, the spectra of high-mass X-ray binaries are dominated by extremely hot accretion disks and peak at hard X-ray energies in the 1--10 keV range (Figure~\ref{fig:sedcomp}).
Their spectra can be approximated with a multi-color disk (MCD) model \citep{mitsudaEnergySpectraLowmass1984a}, which produces a modified blackbody spectrum representing gas at a range of temperatures in the accretion disk.
This model is parameterized by the mass of the accreting black hole \mbh{} and the maximum radius of the disk \rmax{} \citep[in this work, we compute this spectrum using code from the \textsc{ARES} package\footnote{\url{https://bitbucket.org/mirochaj/ares}};][]{mirochaDecodingXrayProperties2014}.
For black hole masses in the range of 10--100 $M_\odot$ and radii of $10^3$--$10^5$ cm, 54.4 eV remains solidly in the Rayleigh-Jeans tail of this spectrum, with flux far lower than at the SED peak in the keV range.
Over this range, the black hole mass has the largest impact on flux at 54.4 eV for a fixed total luminosity, with more massive black holes yielding a softer spectrum and more flux at 54.4 eV.
Increasing \rmax{} from $10^3$ to $10^4$ cm increases the flux at 54.4 eV by $\lesssim 25\%$ for black hole masses in this range, and essentially no change is seen when \rmax{} is increased further to $10^5$ cm.
Accordingly, we fix $\rmax{}=10^4$ cm for our grid.
HMXBs are known to undergo spectral transitions, though generally their spectrum is found to harden relative to an MCD spectrum likely due to Compton up-scattering during a super-Eddington accretion event \citep[e.g.][]{kaaretStateTransitionLuminous2013,brorbyTransitionXrayBinary2015}, decreasing the flux at low ($\ll 1$ keV) energies.
A more sophisticated model including Comptonization would thus cause the HMXB spectrum to be less efficient at producing photons at the \heplus{}-ionizing edge.
For simplicity in this analysis, we adopt a single MCD model to represent the HMXB contribution to the total SED.
While in massive galaxies we expect the X-ray spectrum to be a composite of several active HMXBs, the total unresolved X-ray luminosity of most nearby dwarf galaxies studied in \citet[][$10^{38}$--$10^{40}$ erg/s]{brorbyXrayBinaryFormation2014,brorbyEnhancedXrayEmission2016} is actually suggestive of the luminosity of individual HMXBs resolved in larger spiral galaxies \cite[e.g.][]{mineoXrayEmissionStarforming2012}, so this assumption may actually be more appropriate for the study of \heii{} in the lowest-metallicity, lowest-mass galaxies found locally.

We are interested in the impact of explicitly varying the HMXB contribution on high-ionization nebular emission.
Thus, we incorporate as an additional parameter the X-ray luminosity measured in a broad \chandra{} band (erg/s) produced per unit of star formation ($\mathrm{M_\odot/yr}$), hereafter referred to as the X-ray production efficiency: \lxb{}/SFR.
Previous observations with \chandra{} provide guidance as to the range of values of this parameter attained in local galaxies.
For the purposes of this study, we allow this quantity to vary over a broad range bracketing the values typically measured in nearby dwarf galaxies \citep[e.g.][]{brorbyEnhancedXrayEmission2016} by several orders of magnitude: $10^{40}$--$10^{44}\mathrm{erg/s/(M_\odot/yr)}$.

Figure~\ref{fig:sedcomp} compares these two SED components for a representative model in our grid.
We plot here a BPASS model computed with the above assumptions (constant star formation history at 1 $\mathrm{M_\odot/yr}$ and fiducial IMF) at $Z=0.004$ ($Z/Z_\odot=0.2$).
Next to this, we plot an MCD spectrum assuming $\rmax{}=10^4$ cm for $\mbh{}=10$--100 $\rm M_\odot$, in each case normalized in the 0.5--8 keV band to the median X-ray production efficiency in our grid: $10^{42} \mathrm{erg/s/(M_\odot/yr)}$.
This particular X-ray production efficiency is chosen such that the HMXB spectrum begins to impact significantly on the total EUV flux.
At this X-ray luminosity per unit of star formation, the Rayleigh-Jeans tail of the MCD spectrum approaches the stellar contribution to the SED at the \heplus{}-ionizing edge as the black hole mass is increased.

\subsubsection{Photoionization Modeling}

Accurately predicting the nebular emission powered by HMXBs requires photoionization modeling, taking into account the joint effects of both HMXBs and stars in heating and ionizing their surrounding gas.
We use the code \cloudy{} \citep[version 17.1, detailed in][]{ferland2017ReleaseCloudy2017} to perform this part of the analysis.
This photoionization code natively handles the processing of X-rays by nebular gas, including accounting for secondary ionizations.
Following the standard methodology for modeling integrated light from star-forming galaxies \citep[e.g.][]{bylerNebularContinuumLine2017}, we assume a closed radiation-bounded spherical geometry consisting of a shell of material located sufficiently far from the central source ($10^{19}$ cm) to be essentially plane-parallel.
We allow the computation to iterate to convergence up to 5 times (typically only 3 are required), and stop the calculation once the temperature drops below 100 K or the edge of the hydrogen ionization front is reached (beyond either of which nebular line emission will be minimal).
We run the \cloudy{} models with a separate high-resolution input file for each SED, and measure the strength of the emission lines we are interested in directly from the saved output.
To organize and analyze the grid, we utilize a modified version of the grid construction tools developed as part of \textsc{cloudyfsps} \citep{nellbylerNellbylerCloudyfspsInitial2018}.
Our approach is appropriate for our model of radiation-bounded \hii{} regions surrounding recently-formed young stellar populations, but modeling more massive galaxies with several dominant generations of stars generally benefits from more complex approaches \citep[e.g.][]{charlotNebularEmissionStarforming2001}.
Considering density-bounded model conditions is also beyond the scope of this paper, but \citet{platConstraintsProductionEscape2019} demonstrate that the suppression of lower-ionization emission that these conditions introduce is generally inconsistent with the nebular properties of a broader sample of local \heii{}-emitters.

There are several other important variables describing the gas to consider in this modeling scheme.
We fix the gas density to $n_\mathrm{H}=10^2$ \unit{cm^{-2}} to approximate the typical values measured from [\ion{S}{ii}] in nearby star-forming galaxies, including those which power strong high-ionization line emission \citep[e.g.][]{brinchmannNewInsightsStellar2008,senchynaUltravioletSpectraExtreme2017,bergChemicalEvolutionCarbon2019a}.
While typical \hii{} region densities may be somewhat higher in galaxies in the early universe \citep[e.g.][]{shiraziStarsWereBorn2014,sandersMOSDEFSurveyElectron2016}, this is unlikely to significantly affect recombination lines such as \heii{} \citep[e.g.][]{platConstraintsProductionEscape2019}.
Since in this work we focus on metal-poor star-forming dwarf galaxies with low dust extinction measurements, we do not consider the effects of extinction or depletion onto dust in this analysis \citep[though these effects are very important for general galaxy modeling; see e.g.][]{gutkinModellingNebularEmission2016}.
The attenuation introduced by dust extinction is minimal for X-ray photons \citep[dominated by small-angle forward scattering;][]{drainePhysicsInterstellarIntergalactic2011} and we only consider observed optical line ratios after correction for extinction according to the Balmer decrement (Section~\ref{sec:obs}).

Next, we must consider the gas-phase metallicity and chemical abundance patterns, which directly impact upon both metal line strengths and thereby the gas cooling efficiency and thus temperature.
We fix the gas abundance patterns to those adopted by \citet{dopitaTheoreticalRecalibrationExtragalactic2000}, which is based upon the solar abundances found by \citet{andersAbundancesElementsMeteoritic1989} with an additional empirically-motivated scaling of $\mathrm{N/O}$ with $\mathrm{O/H}$ imposed to account for the secondary production of nitrogen.
In this first analysis, we scale the gas-phase metal abundances directly with the metallicity $Z$ of the stellar population, while noting that allowing for an offset in this ratio could enhance the contribution of stellar ionizing flux to the production of high-ionization metal lines \citep[e.g.][]{steidelReconcilingStellarNebular2016,senchynaExtremelyMetalpoorGalaxies2019}.

Finally, we also allow the gas ionization parameter $U$ to vary.
This dimensionless quantity parametrizes the relative density of hydrogen-ionizing photons to the gas density:
\begin{equation} 
    U = \frac{Q_\mathrm{H}}{4\pi R^2 n_\mathrm{H} c} ,
\end{equation}
where $Q_\mathrm{H}$ is the total number of hydrogen-ionizing photons emitted by the source spectrum per second, $c$ is the speed of light, and $R$ is the radius of the ionized region.
In this effectively plane-parallel case, the Str\"{o}mgren sphere radius is approximately equal to the inner radius of the shell, and their distinction is unimportant \citep[e.g.][]{charlotNebularEmissionStarforming2001,bylerNebularContinuumLine2017}.
Nebular line ratios are typically insensitive to changes in $R$ and $Q_\mathrm{H}$ that preserve $\log U$  \citep[e.g.][]{evansTheoreticalModelsII1985,mccallChemistryGalaxiesNature1985}.
Since we fix the radius, specifying the ionization parameter implicitly normalizes the input spectrum.
We allow this parameter to vary between $-3<\log U < -1$, which spans the range typically observed in galaxies dominated by recent star formation.

\subsection{Modeling Results and Predictions}
\label{subsec:modelres}

The framework described above results in a model for a composite stellar and HMXB spectrum with four free parameters.
These are metallicity $Z$ (of both the stars and the gas, which we couple); the ionization parameter of the gas $\log U$; the X-ray production efficiency, $\lxb/\mathrm{SFR}$; and the black hole mass for our MCD model, $M_\mathrm{BH}$.
We construct a grid of models while varying these parameters, and extract the CLOUDY predicted line fluxes for each point therein, allowing us to investigate the relative impact of these variables on nebular line emission.

\begin{figure}
    \includegraphics[width=0.5\textwidth]{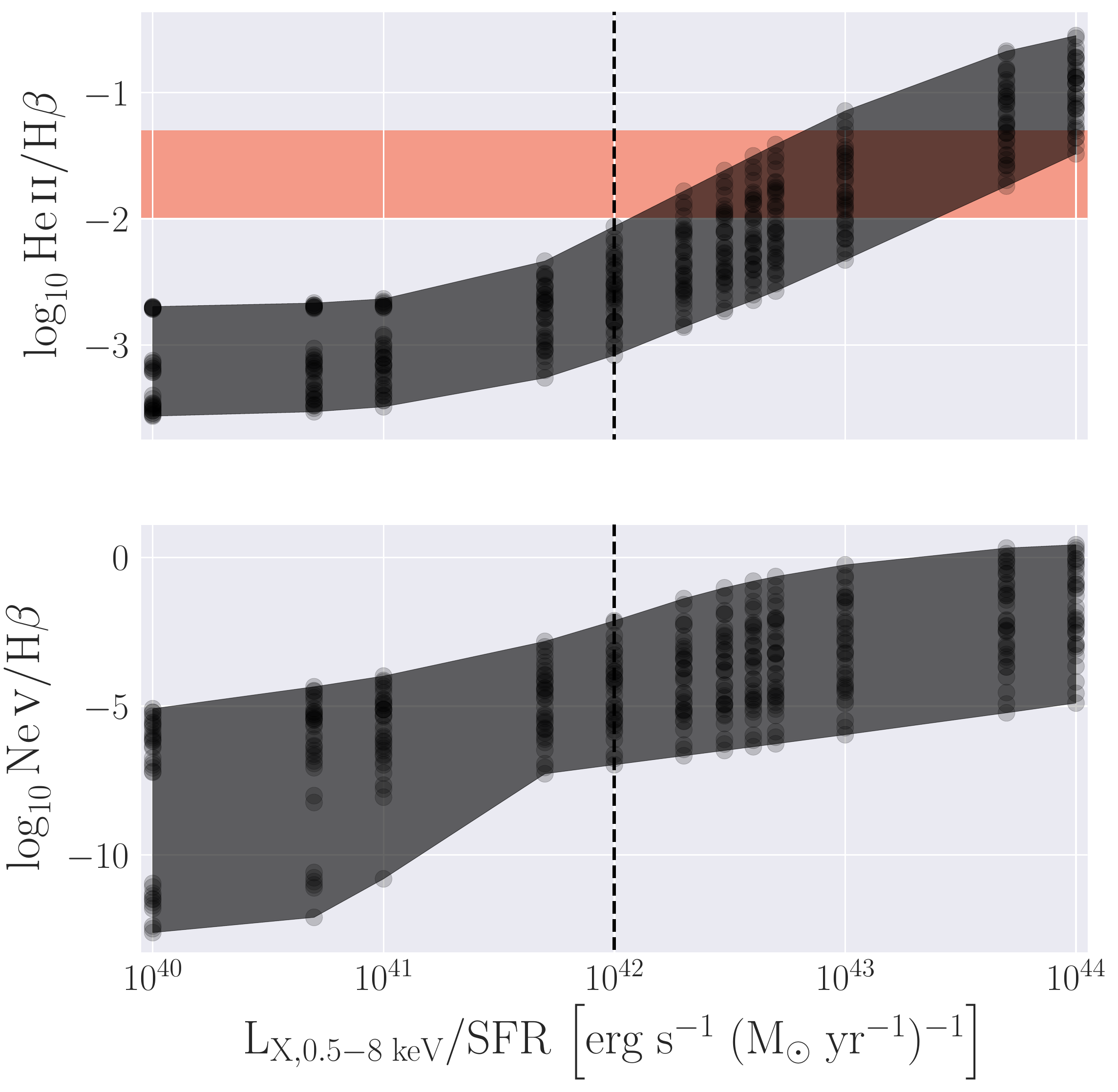}
    \caption{
        Line fluxes for \heii{} $\lambda 4686$ and \nev{} $\lambda 3426$ (top and bottom panels, respectively) relative to H$\beta$ for all points in our model grid, plotted as a function of the X-ray luminosity per unit of star formation.
        The individual photoionization model grid points are displayed as black circles, and we outline in grey the flux range spanned by the models at each value of $\lxb/\mathrm{SFR}$.
        In the top panel for \heii{}, we shade in red the range of values observed for this line in low-metallicity nearby star-forming galaxies \citep[e.g.][]{shiraziStronglyStarForming2012,senchynaUltravioletSpectraExtreme2017,senchynaExtremelyMetalpoorGalaxies2019}.
        \heii{} is only boosted into the observed range of values for this line by the addition of the HMXB spectrum for systems with very high X-ray production efficiencies ($\lxb{}/\mathrm{SFR}>10^{42}\, \unit{erg\, s^{-1} / (M_\odot \, yr^{-1})}$, marked by a vertical black dashed line).
    }
    \label{fig:heiinevnv_all}
\end{figure}

We are particularly interested in how varying the HMXB contribution to the spectrum impacts the high-ionization nebular lines.
In Figure~\ref{fig:heiinevnv_all}, we plot the predicted line fluxes for \heii{} $\lambda 4686$ and \nev{} $\lambda 3426$ relative to H$\beta$ for our entire grid as a function of the X-ray production efficiency.
These two species have ionization potentials of 54 eV and 97 eV (respectively).
As the X-ray production efficiency is increased from $10^{40}$--$10^{44}$ \unit{erg/s} at a star formation rate of 1 \unit{M_\odot/year}, we find that the range of flux in \heii{} relative to H$\beta$ is increased by 2 orders of magnitude.
The effect on \nev{} is even more dramatic, increasing from essentially unobservable at $<10^{-5}$ times the flux of H$\beta$ to uniformly greater than this limit for the highest X-ray productions rates.
At the highest X-ray luminosities per SFR tested, both lines approach the flux of H$\beta$, indicating that the composite spectrum is effectively dominated by the hard Rayleigh-Jeans tail of the HMXB model.
If we assume HMXBs provide the photons necessary to power \heii{} and \nev{}, then our models suggest we should observe a trend between enhanced flux in these lines relative to H$\beta$ and the X-ray production efficiency.

However, \heii{} and \nev{} are only substantially affected by the HMXB spectrum for systems with very large X-ray output relative to their star formation rate.
As the X-ray production efficiency is increased from $10^{40}$ to $10^{41}$ erg/s per unit SFR, the ratio of \heii{} $\lambda 4686$/H$\beta$ remains in the range observed for the BPASSv2.2 models alone, with the maximum achieved value increasing only from $\heii{}/\mathrm{H}\beta = 0.20 \%$ to $0.23\%$.
This is an order of magnitude below measurements in metal-poor dwarf galaxies, which span 1--5\% \citep[the red band in Figure~\ref{fig:heiinevnv_all}, e.g.][]{shiraziStronglyStarForming2012,senchynaExtremelyMetalpoorGalaxies2019}.
None of the models exceed 1\% flux in \heii{} or \nev{} relative to H$\beta$ until $\lxb/\mathrm{SFR}$ exceeds $10^{42} \, \unit{erg\, s^{-1}}$ at $1 \; \mathrm{M_\odot \, yr^{-1}}$ (marked with a dashed black line in Figure~\ref{fig:heiinevnv_all}).
That is, galaxies with HMXB populations less luminous per SFR than this cutoff do not produce sufficient flux in the EUV to power \heii{} at the level observed locally.

\begin{figure}
    \includegraphics[width=0.5\textwidth]{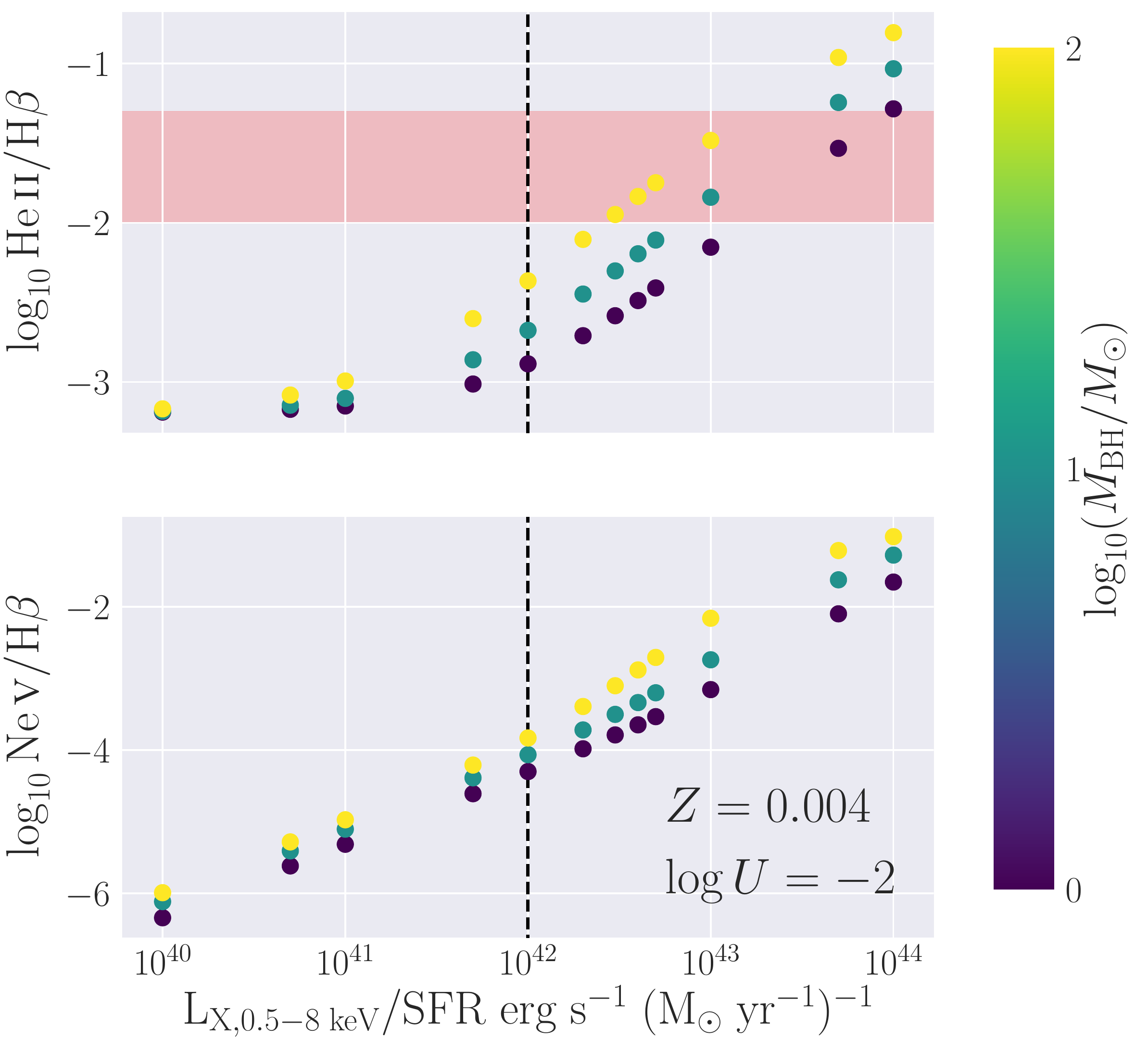}
    \caption{
        Same as Figure~\ref{fig:heiinevnv_all}, but displaying a subset of the models with metallicity and ionization parameter fixed to $Z=0.004$ and $\log U=-2$ to illustrate the dependence of the nebular lines on $M_\mathrm{BH}$.
        Increasing the black hole mass from 1--100 $M_\odot$ can effectively boost the prominence of \heii{} by up to nearly an order of magnitude, but this only has a modest effect at low $\lxb/\mathrm{SFR}$ values where the the stellar spectrum dominates the EUV.
    }
    \label{fig:heiinevnv_bhmass}
\end{figure}

\begin{figure}
    \includegraphics[width=0.5\textwidth]{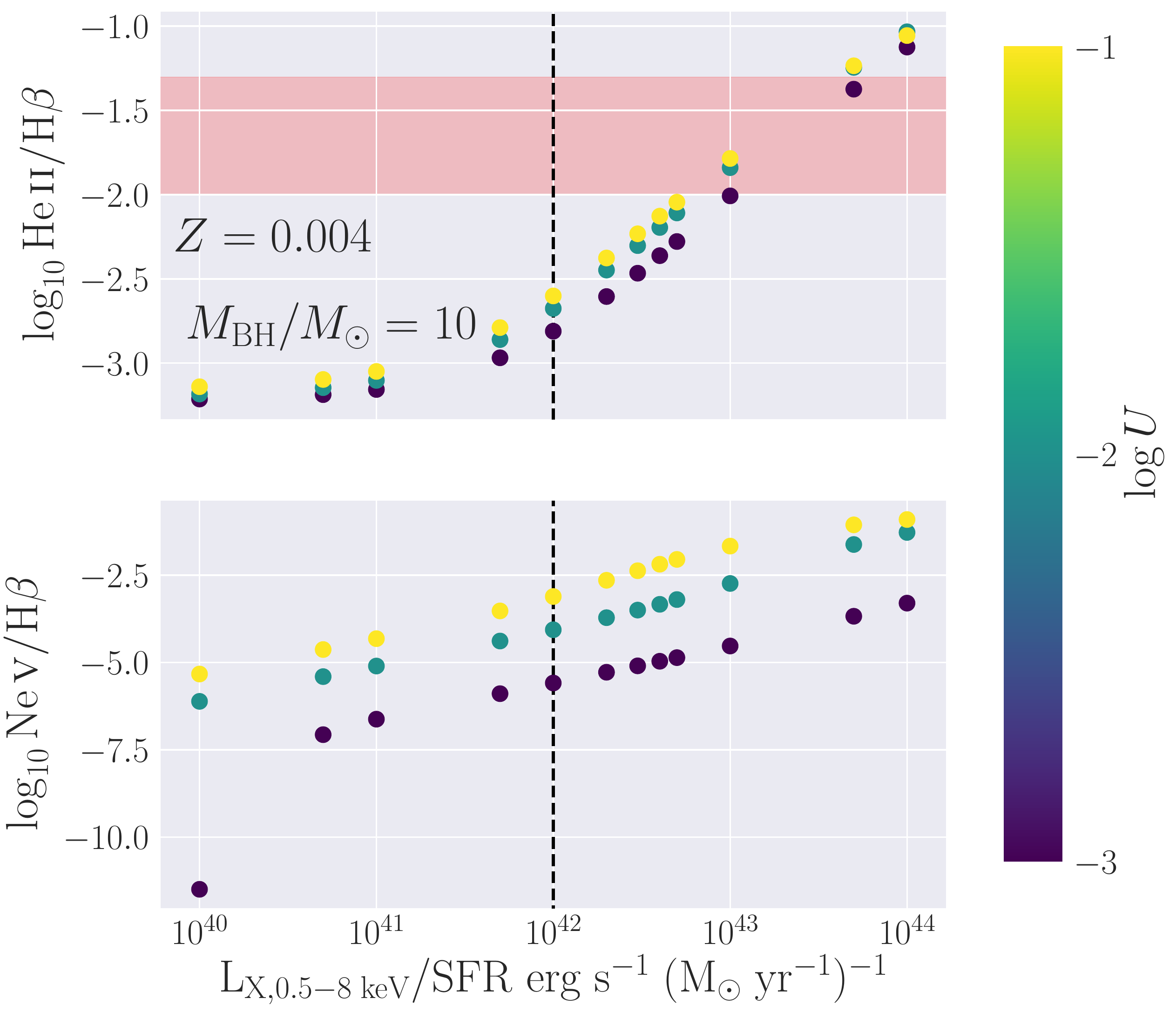}
    \caption{
        Same as Figure~\ref{fig:heiinevnv_all}, but with metallicity and black hole mass fixed to $Z=0.004$ and $M_\mathrm{BH}/M_\odot=10$ to investigate the dependence of the various lines on $\log U$.
        The relative strength of \nev{} to H$\beta$ is strongly dependent on $\log U$, increasing with this quantity by up to more than five orders of magnitude with other parameters fixed.
        However, \heii{} is only modestly affected by an increase in the ionization parameter (especially at low \lxb{}), due to the fact that both \heii{} and H$\beta$ are recombination lines.
    }
    \label{fig:heiinevnv_logu}
\end{figure}

\begin{figure}
    \includegraphics[width=0.5\textwidth]{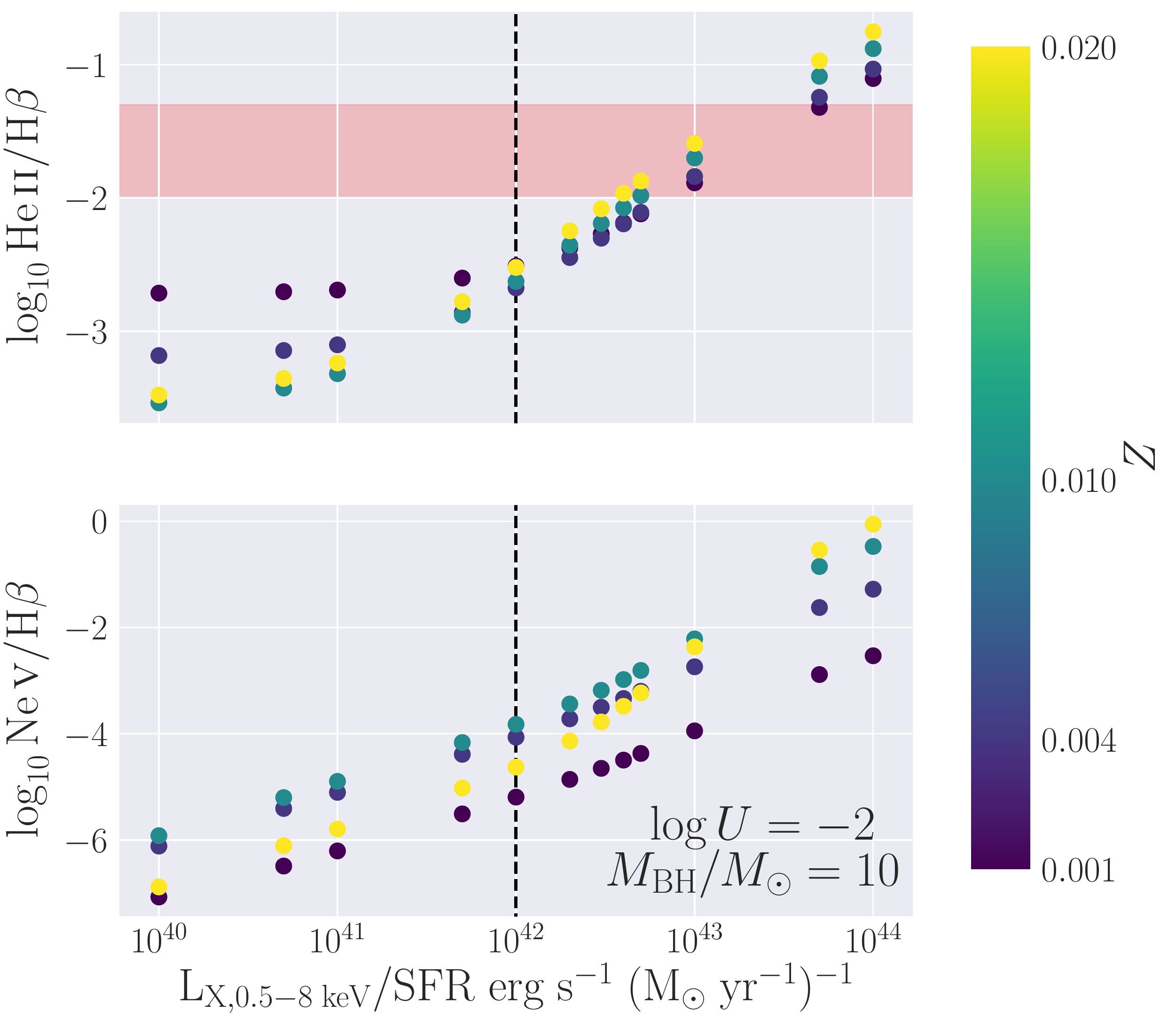}
    \caption{
        Same as Figure~\ref{fig:heiinevnv_all}, but with the ionization parameter and black hole mass fixed to $\log U=-2$ and $M_\mathrm{BH}/M_\odot=10$, while the metallicity $Z$ is allowed to vary.
        At low X-ray production efficiencies where the stellar SED dominates (left of the dashed line), \heii{} is substantially more affected by decreasing metallicity than by $\log U$ or $M_\mathrm{BH}$, increasing by nearly an order of magnitude as the metallicity is decreased.
        \nev{} shows a more complicated behavior, due to the competing effects of increased gas temperature and diminishing neon abundances with decreasing metallicity.
    }
    \label{fig:heiinevnv_met}
\end{figure}

Though both \heii{} and \nev{} increase in strength relative to H$\beta$ as the X-ray production efficiency is increased, the dispersion in their flux at fixed \lxb{}/SFR evident in Figure~\ref{fig:heiinevnv_all} illustrates that they are also sensitive to the other model parameters ($M_\mathrm{BH}$, $\log U$, and $Z$).
We now investigate the impact of these variables in-turn, starting with the model black hole mass.
In Figure~\ref{fig:heiinevnv_bhmass}, we fix the metallicity and ionization parameter to $Z=0.004$ and $\log U=-2$ while allowing $M_\mathrm{BH}$ to vary over our full grid range (1--100 $M_\odot$).
At X-ray production efficiencies below $10^{42}$ erg/s at 1 \unit{M_\odot/year}, the modeled black hole mass has essentially no effect, since the EUV is still largely dominated by the stellar SED.
At higher values of \lxb{}/SFR where the HMXB spectrum begins to impact significantly on the EUV, we see that increasing the black hole mass boosts both \nev{} and especially \heii{} (since it is produced at lower energies).
This is because a higher black hole mass yields a softer MCD spectrum (Figure~\ref{fig:sedcomp}), allowing the HMXB flux to play a larger role in the ionization structure of the nebula.

The gas geometry can also significantly affect the prominence of \heii{} and \nev{} with the input spectrum fixed (Figure~\ref{fig:heiinevnv_logu}).
Increasing $\log U$ from $-3$ to $-1$ can increase $\nev{}/\mathrm{H\beta}$ by nearly 3 orders of magnitude, by effectively expanding the size of the highest-ionization part of the \hii{} region.
However, $\log U$ has a smaller effect on the flux of \heii{} relative to H$\beta$ ($\lesssim 0.2$ dex).
Since \heii{} and H$\beta$ are recombination lines, their ratio is fixed to first-order by the hardness of the input ionizing spectrum \citep[e.g.][]{hummerRecombinationlineIntensitiesHydrogenic1987,drainePhysicsInterstellarIntergalactic2011,senchynaUltravioletSpectraExtreme2017}.

The metallicity also plays a role in modulating the strength of \heii{} and \nev{}.
In the regime where the stellar SED dominates \heii{} production (left of the dashed line in Figure~\ref{fig:heiinevnv_met}), \heii{} is most strongly affected by metallicity, increasing by nearly an order of magnitude as metallicity drops from $Z=0.02$ to $0.001$.
As mentioned in Section~\ref{subsec:modelmeth}, this is due to the hardening of the stellar SED with decreasing metallicity.
At high X-ray production efficiencies (right of the dashed line), \heii{}/H$\beta$ decreases modestly with decreasing metallicity, due to the lowering of the recombination rate with increasing gas temperature \citep{hummerRecombinationlineIntensitiesHydrogenic1987}.
In contrast, \nev{}/H$\beta$ displays a very different behavior; at all but the highest X-ray production efficiencies, \nev{}/H$\beta$ peaks at $Z=0.01$ before dropping at lower metallicity.
Since the adopted stellar SED is too soft to contribute effectively to these lines, this behavior is due primarily to gas physics, with a competition between the diminishing abundance of neon and the increasing electron temperature (due to less efficient cooling).
At the lowest metallicities ($Z<0.01$), the diminishing neon abundance dominates, and \nev{} become less prominent with decreasing metallicity.
At the highest X-ray production efficiencies, gas heating is far more efficient, and this line simply decreases in strength with decreasing abundance over the entire metallicity range.

While $M_\mathrm{BH}$, $\log U$, and $Z$ all impact on \heii{} and \nev{} to some degree, our full set of photoionization models suggest two testable observational predictions independent of these variables if we assume that HMXBs dominate \heii{} production.
First, strong \heii{} emission should be associated with very high X-ray production efficiencies: that is, X-ray luminosities in-excess of $10^{42}$ erg/s for a galaxy forming stars at 1 \unit{M_\odot/year}.
And second, stronger \heii{} relative to H$\beta$ (and, with greater dispersion due to gas physics, \nev{}/H$\beta$) should positively correlate with higher X-ray luminosities per unit star formation rate.
In the remainder of this paper, we will test whether these predictions hold in a sample of local star-forming galaxies with both high-quality \chandra{} and optical line constraints.

\section{Observations}
\label{sec:obs}

Through photoionization modeling, we have established an observational test of the scenario in which HMXBs dominate the production of \heii{} in star-forming galaxies (Section~\ref{sec:model}).
To investigate this, we have collated archival \chandra{} ACIS-S imaging of nearby $z<0.2$ star-forming dwarf galaxies and assembled both ESI and new MMT spectra covering \heii{} $\lambda 4686$ and (in several cases) \nev{} $\lambda 3426$.
Though X-ray detections have previously been published for several of the targets in our sample, the absence of reported coordinates for these X-ray sources and the fact that several of our collected datasets are unpublished requires that we reanalyze these X-ray data.
We present the sample in Sections~\ref{sec:sample} and \ref{subsec:basicprop}; the optical spectroscopy in Section~\ref{sec:optspec}; the \chandra{} data in Section~\ref{sec:chandra}; and describe the results of this joint optical and X-ray analysis in Section~\ref{subsec:casebycaseres}.

\subsection{Selection of the Sample}
\label{sec:sample}

In order to test whether HMXBs are responsible for \heii{}, we need an observational sample with high-quality constraints on both X-ray emission and nebular \heii{}.
In particular, the X-ray imaging must have sufficiently high spatial resolution ($\lesssim 1\arcsec$) to confidently associate X-ray point sources with the spectroscopic targets, which is presently only attained by \chandra{}.
In addition, the optical spectra must have sufficiently high spectral resolution and signal-to-noise to confidently deblend the narrow ($<500$ km/s) nebular \heii{}$\lambda 4686$ recombination line from the broad ($>1000$ km/s) component originating in the winds of WR or very massive O stars, which is generally not possible with SDSS spectra \citep[see e.g.][]{senchynaUltravioletSpectraExtreme2017}.
In the past several years, we have acquired high resolution optical spectra for a total of eleven star-forming dwarf galaxies which have \chandra{} X-ray imaging observations available in the archive (Table~\ref{tab:obssumm}).
Though selected in different ways initially, all are actively star-forming galaxies at subsolar metallicity and thus potential hosts for metal-poor HMXB populations.

Eight of these eleven are galaxies targeted with \hstcos{} ultraviolet spectroscopy in \citet{senchynaUltravioletSpectraExtreme2017} and \citet{senchynaExtremelyMetalpoorGalaxies2019} for which we also obtained deep echellette optical spectra (we will subsequently refer to these papers as \citetalias{senchynaUltravioletSpectraExtreme2017} and \citetalias{senchynaExtremelyMetalpoorGalaxies2019}, respectively).
These galaxies were originally selected on the basis of either a detection of optical \heii{} emission in an SDSS spectrum \citepalias{senchynaUltravioletSpectraExtreme2017} and as having gas-phase metallicity measurements placing them in the regime of extremely metal-poor galaxies \citepalias[XMPs, $12+\log\mathrm{O/H}<7.7$, $Z/Z_\odot \lesssim 0.1$;][]{senchynaExtremelyMetalpoorGalaxies2019}.
All are compact regions of active star formation dominated by the light from young (typically $<50$ Myr) stellar populations.
We cross-matched the entire sample of sixteen presented in these publications with publicly-available observations in the \chandra{} Observation Catalog accessed via ChaSeR to obtain the subsample of eight with \chandra{} coverage discussed here\footnote{Note that while SB198 from \citetalias{senchynaUltravioletSpectraExtreme2017} falls within the standard 10\arcmin{} search radius employed by ChaSeR, the optical target falls outside the footprint of the ACIS-S3 chip.}.
We refer the reader to these previous papers for a more thorough discussion of the sample properties.

We also obtained new deep optical spectra for three additional galaxies previously studied with \chandra{}.
In particular, we targeted three galaxies classified as Lyman Break Analogues \citep[LBAs;][]{heckmanPropertiesUltravioletluminousGalaxies2005} and studied in X-rays using \chandra{} by \citet{basu-zychEvidenceElevatedXRay2013} and \citet[][hereafter \citetalias{brorbyEnhancedXrayEmission2016}]{brorbyEnhancedXrayEmission2016} with the MMT Blue Channel spectrograph: SHOC042, J2251+1427, and SHOC595.
These spectra provide coverage of both \heii{} $\lambda 4686$ and \nev{} $\lambda 3426$.
We will discuss the reduction and analysis of these observations in Section~\ref{sec:optspec} below.

\subsection{Star Formation Rates and General Characterization}
\label{subsec:basicprop}

\begin{table*}
    \centering
    \caption{\label{tab:obssumm} Basic properties of the eleven galaxies analyzed in this work, ordered by right ascension.
    Redshifts are measured from the strong optical lines in SDSS spectra \citep[or, in the case of two of the XMPs, from an MMT spectrum;][]{senchynaExtremelyMetalpoorGalaxies2019}, and the corresponding distances computed using the local velocity flow model presented by \citet{tonrySurfaceBrightnessFluctuation2000} as described in \citep{senchynaUltravioletSpectraExtreme2017,senchynaExtremelyMetalpoorGalaxies2019}. 
    We also present gas-phase metallicities measured using the direct-$T_e$ method and star formation rates measured using the Balmer lines.
    }

\begin{tabular}{lccccccccc}
\hline
Name & RA & Dec & Sample & z & Distance & $12+\log\mathrm{O/H}$ & $W_{0,\mathrm{H\beta}}$ & $\log_{10} \mathrm{SFR(H\beta)} / (M_\odot/\mathrm{yr})$\\ 
 & (J2000) & (J2000) &  &  & (Mpc) & (direct-$T_e$) & (\AA{}) & \\ 
\hline
SHOC042& 00:55:27.46& -00:21:48.7& \citetalias{brorbyEnhancedXrayEmission2016}& 0.1674& 805& $8.21 \pm 0.09$& $47 \pm 4$& $0.77 \pm 0.03$\\ 
J0940+2935& 09:40:12.87& +29:35:30.2& \citetalias{senchynaExtremelyMetalpoorGalaxies2019}& 0.0024& 8& $7.63 \pm 0.14$& $37 \pm 1$& $-3.58 \pm 0.10$\\ 
SB80& 09:42:56.74& +09:28:16.2& \citetalias{senchynaUltravioletSpectraExtreme2017}& 0.0109& 46& $8.24 \pm 0.06$& $243 \pm 17$& $-0.90 \pm 0.02$\\ 
SB2& 09:44:01.87& -00:38:32.2& \citetalias{senchynaUltravioletSpectraExtreme2017}& 0.0048& 19& $7.81 \pm 0.07$& $274 \pm 17$& $-1.61 \pm 0.02$\\ 
J1119+5130& 11:19:34.37& +51:30:12.0& \citetalias{senchynaExtremelyMetalpoorGalaxies2019}& 0.0045& 22& $7.51 \pm 0.07$& $50 \pm 1$& $-2.31 \pm 0.04$\\ 
SBSG1129+576& 11:32:02.64& +57:22:36.4& \citetalias{senchynaExtremelyMetalpoorGalaxies2019}& 0.0050& 25& $7.47 \pm 0.06$& $81 \pm 3$& $-2.45 \pm 0.05$\\ 
SB191& 12:15:18.60& +20:38:26.7& \citetalias{senchynaUltravioletSpectraExtreme2017}& 0.0028& 10& $8.30 \pm 0.07$& $392 \pm 23$& $-2.20 \pm 0.02$\\ 
SB111& 12:30:48.60& +12:02:42.8& \citetalias{senchynaUltravioletSpectraExtreme2017}& 0.0042& 16& $7.81 \pm 0.08$& $102 \pm 5$& $-2.37 \pm 0.02$\\ 
HS1442+4250& 14:44:11.46& +42:37:35.6& \citetalias{senchynaExtremelyMetalpoorGalaxies2019}& 0.0023& 11& $7.65 \pm 0.04$& $113 \pm 4$& $-3.03 \pm 0.05$\\ 
J2251+1327& 22:51:40.32& +13:27:13.4& \citetalias{brorbyEnhancedXrayEmission2016}& 0.0621& 279& $8.29 \pm 0.06$& $56 \pm 1$& $0.22 \pm 0.02$\\ 
SHOC595& 23:07:03.76& +01:13:11.2& \citetalias{brorbyEnhancedXrayEmission2016}& 0.1258& 589& $8.07 \pm 0.10$& $61 \pm 4$& $0.49 \pm 0.04$\\ 
\hline
\end{tabular}

\end{table*}

Before advancing to the high-quality optical spectra and \chandra{} data available for the galaxies in our sample, we will first discuss their bulk properties.
All but one of the targeted systems (HS1442+4250) has an SDSS spectrum available as of the latest data release \citep[DR15;][]{aguadoFifteenthDataRelease2019}, and the other (HS1442+4250) has an MMT spectrum described in \citetalias{senchynaExtremelyMetalpoorGalaxies2019}.
We derive distance estimates by comparing the redshifts measured in these spectra to the local flow model presented by \citet{tonrySurfaceBrightnessFluctuation2000} with $H_0=70$ km/s/Mpc, with this simply reverting to the cosmological distances for the most distant objects.
We present these distances in Table~\ref{tab:obssumm}, along with gas-phase metallicities and the references to their derivation.

Since the higher-resolution spectra discussed in Section~\ref{sec:optspec} do not all provide access to both H$\beta$ and H$\alpha$, we obtain uniformly-calibrated measurements of the strong optical lines from the SDSS spectra for reddening correction and SFR estimation.
We follow the procedure outlined in \citetalias{senchynaUltravioletSpectraExtreme2017,senchynaExtremelyMetalpoorGalaxies2019}: in particular, we measure line fluxes using a custom \textsc{python} routine based on MCMC fits \citep{foreman-mackeyEmceeMCMCHammer2013} with a linear continuum plus Gaussian line model, and correct for extinction by comparing the observed Balmer decrement to that predicted from our direct determination of $T_e$ with \textsc{PyNeb} \citep{luridianaPyNebNewTool2015}, assuming an SMC extinction curve \citep{gordonQuantitativeComparisonSmall2003} after correction for Galactic extinction \citep{schlaflyMeasuringReddeningSloan2011,fitzpatrickCorrectingEffectsInterstellar1999}.
We follow the procedure described in \citetalias{senchynaUltravioletSpectraExtreme2017} using \textsc{PyNeb} to derive direct temperature metallicities (with [\oiii{}]$\lambda 4363$) for the three LBAs from the \citetalias{brorbyEnhancedXrayEmission2016} sample, yielding uniform metallicity measurements for the entire sample (Table~\ref{tab:obssumm}).
In the case of HS1442+4250, we use an MMT Blue Channel spectrum obtained with the 300 lines/mm grating on January 25, 2017 \citepalias[described in][]{senchynaExtremelyMetalpoorGalaxies2019}.
To correct for aperture differences between this MMT spectrum and the other SDSS spectra, we derive an effective aperture correction by comparing the measured fluxes of the strong lines of H and O in MMT spectra taken in the same configuration and on the same night for SBSG1129+576 and J1119+5130 to their SDSS spectra.
This yields consistent correction factors of $0.58$ and $0.57$ for SBSG1129+576 and J1119+5130 with a scatter of $\sim 3\%$ among the different lines, respectively; we adopt a value of $0.57$ and add an additional statistical uncertainty of 10\% in this conversion.

These measurements of H$\alpha$ and H$\beta$ provide a direct estimate of the current star formation rate in the spectral aperture where \heii{} is constrained.
Following the standard methodology, we convert the dust-corrected Balmer line luminosities into SFR estimates using a conversion factor derived from the same BPASS v2.2 models employed in the photoionization modeling (Section~\ref{sec:model}).
In particular, we assume $Z=0.003$ (corresponding to the mean metallicity of our sample, $12+\log\mathrm{O/H}=7.9$, assuming solar abundances) and the fiducial BPASS IMF (Salpeter slope).
For each galaxy, we derive an SFR conversion factor appropriate for the effective age of the dominant stellar population by choosing the constant star formation rate model with predicted H$\beta$ equivalent width closest to that observed.
The uncertainty in the resulting estimates (Table~\ref{tab:obssumm}) represents the statistical uncertainties in the flux measurement and determination of $Q(\mathrm{Lyc})/L_{\mathrm{H}\alpha}$.
These estimates only account for star formation within the spectroscopic aperture, and thus represent lower limits to the total star formation rate of the target galaxies.

These bulk measurements reveal a diverse set of actively star-forming galaxies.
They range in distance from $\sim 10$ to 800 Mpc, with all but the relatively distant LBAs from \citetalias{brorbyEnhancedXrayEmission2016} residing at $<50$ Mpc (redshifts $z\lesssim 0.01$).
Their gas-phase metallicities span from the extremely metal-poor regime $12+\log\mathrm{O/H}= 7.5$--$7.7$ up to over half solar metallicity at $12+\log\mathrm{O/H}=8.3$.
Though they are all dominated by young stars, their absolute star formation rates span over four orders of magnitude, from $10^{-3.6}$ to $10^{0.77}$ \unit{M_\odot/year}.

\subsection{Optical Spectroscopy}
\label{sec:optspec}

Here, we collate and summarize the deep optical spectral constraints available for the systems under study.
As demonstrated in \citetalias{senchynaUltravioletSpectraExtreme2017}, nebular and broader stellar \heii{} can be confused at the signal-to-noise and resolution typically attained in SDSS fiber spectra.
We have specifically selected galaxies for which we have additional optical spectral constraints that can separate nebular from stellar \heii{} and in some cases provide constraints on \nev{} $\lambda 3426$, which probes even higher-energy photons.
We obtained optical echelle spectra covering \heii{} and H$\beta$ (but not \nev{}) for eight of the systems with the Echellette Spectrograph and Imager \citep[ESI;][]{sheinisESINewKeck2002} on Keck II in 2016--2017 as part of an ongoing joint optical and ultraviolet spectroscopic campaign \citepalias{senchynaUltravioletSpectraExtreme2017,senchynaExtremelyMetalpoorGalaxies2019}.
We refer the reader to these previous publications for a detailed description of the Keck/ESI observations and analysis, and present the measurements of dust-corrected \heii{}/H$\beta$ derived from these spectra in Table~\ref{tab:optspec}.

We have also obtained several new spectra with the Blue Channel Spectrograph on the 6.5m MMT (MMT/BC) specifically targeting \nev{} and \heii{} (Table~\ref{tab:newmmtspec}).
We observed J2251+1327, SHOC595, and SHOC042 \citepalias[LBAs from][]{brorbyEnhancedXrayEmission2016} on the night of September 10, 2018 with the 800 lines/mm grating and the $1.0''\times 180''$ slit ($0.75$ \AA{}/pixel dispersion), at central wavelengths chosen to span the rest-wavelength range from \nev{} $\lambda 3426$ to H$\beta$ at the SDSS redshift of each target.
The standard stars G24-9, Feige 110, and LB 227 (respectively) and a HeAr/Ne comparison lamp were observed either before or after each target object with the same spectrograph configuration for flux and wavelength calibration.
In addition, on April 15 2018 we used the 300 lines/mm grating to observe an optical point-source identified in SDSS imaging as coincident with an X-ray source in the ACIS-S image of HS1442+4250 (see Section~\ref{sec:datajointres} for more details).
All spectra were reduced using standard longslit techniques with a custom \textsc{python} pipeline.

\begin{table}
    \centering
    \caption{\label{tab:newmmtspec} A summary of the new MMT Blue Channel observations presented in this work. 
    }
    \begin{tabular}{cccc}
        \hline
        Target & Airmass & Configuration & Exposure \\
                &        &  Grating (slit) & (hours) \\
        \hline
        \multicolumn{4}{c}{10 September 2018} \\
        \hline
        \hline
        J2251+1327 & 1.2 & 800gpm ($1.0''\times 180''$) & 2.3 \\
        SHOC595 & 1.3 & 800gpm ($1.0''\times 180''$) & 1.6 \\
        SHOC042 & 1.4 & 800gpm ($1.0''\times 180''$) & 1.6 \\
        \hline
        \multicolumn{4}{c}{15 April 2018} \\
        \hline
        \hline
        \begin{tabular}{@{}l@{}} HS1442+4250 \\X-ray source \end{tabular}  & 1.1 & 300gpm ($1.0''\times 180''$) & 0.17 \\
        \hline
    \end{tabular}
\end{table}

\begin{figure}
    \includegraphics[width=0.5\textwidth]{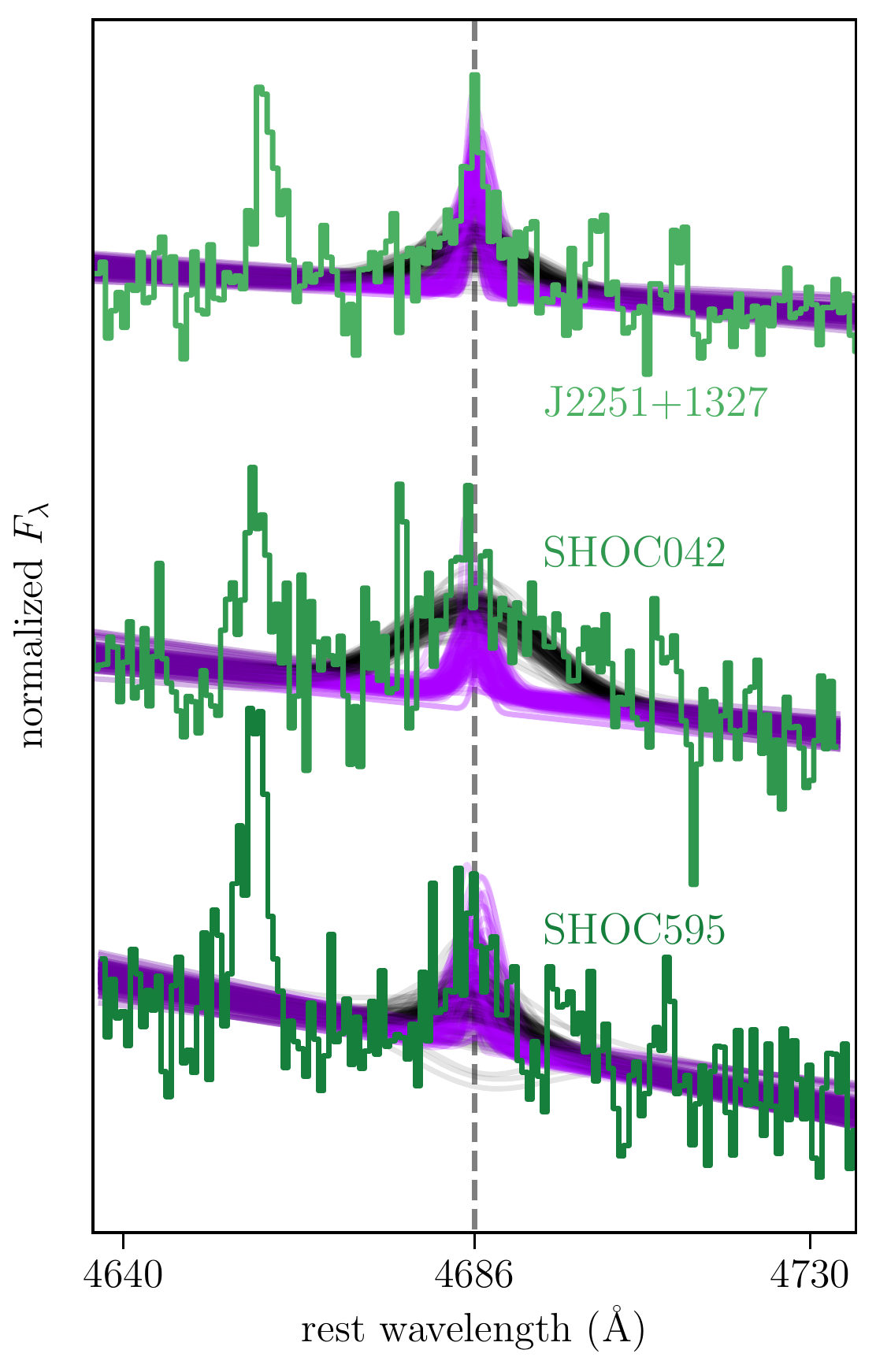}
    \caption{
        The \heii{} $\lambda 4686$ complex for each of the LBAs observed with Blue Channel (green).
        We overplot models drawn from the fit posterior distribution, decomposed into the broad (black) and narrow (purple) components atop the fit linear continuum.
        A narrow component is detected at $>2\sigma$ for J2251+1327 and SHOC042, whereas the profile for SHOC 595 appears to be completely dominated by a line profile with width significantly larger than the strong nebular lines.
    }
    \label{fig:mmtheii}
\end{figure}

To measure nebular \heii{} confidently in these MMT/BC spectra, we follow the same technique adopted with the Keck/ESI data.
In order to distinguish the nebular contribution from the broad $>500$ km/s FWHM stellar component, we fit two Gaussians simultaneously to the \heii{} $\lambda 4686$ line, with one constrained to narrow width comparable to that measured for the strong nebular lines ($\mathrm{FWHM} < 7$ \AA{}) and the other forced to be broader than this cutoff.
The results of these fits are displayed in Figure~\ref{fig:mmtheii}.
We detect a distinct narrow component in J2251+1327 and SHOC042 at a relative strength of \heii{}/H$\beta =0.0070\pm0.0034$ and $0.0046\pm0.0021$ (respectively), and place a $3\sigma$ upper limit on the nebular component of \heii{}/H$\beta<0.0068$ for SHOC595.
These relatively weak detections (\heii{}/H$\beta<0.01$) are consistent with the metallicities of these systems ($12+\log\mathrm{O/H}>8.0$) in the context of the trend towards stronger \heii{} at low metallicity \citep[e.g.\ Figure~7 of][which includes the other systems discussed in this paper]{senchynaExtremelyMetalpoorGalaxies2019}.
In addition, these MMT spectra place strong constraints on the presence of nebular \nev{} $\lambda 3426$ (\nev{}/H$\beta<0.8$ in all three cases).
These line measurements and dereddened $3\sigma$ upper limits where appropriate are reported in Table~\ref{tab:optspec}.
We also include the measurement of \nev{}/H$\beta=0.79\pm0.36$ for SB111 (or J1230+1202) reported by \citet{izotovDetectionNeEmission2012}.
We discuss these measurements in the context of the \chandra{} constraints in the following section.

\begin{table*}
    \centering
    \caption{\label{tab:optspec} Optical spectroscopic constraints on nebular \heii{} and \nev{} for our target galaxies (including $3\sigma$ upper limits where appropriate).
    All measurements have been dereddened using an SMC extinction curve and the Balmer decrement measured in the corresponding SDSS spectrum. 
    }

    \begin{tabular}{ccccc}
        \hline
        Name & \heii{} $\lambda 4686$/H$\beta$ & Source & \nev{} $\lambda 3426$ / H$\beta$ & Source \\
         & $\times 100$ & (\heii{}) & $\times 100$ & (\nev{}) \\
        \hline
        SHOC042 & $0.46\pm0.21$ & MMT/BC (2018-09-10) & $<0.79$ & MMT/BC (2018-09-10) \\
        J0940+2935 & $<1.17$ & Keck/ESI (2017-02-21, \citetalias{senchynaExtremelyMetalpoorGalaxies2019}) &  &  \\
        SB80 & $0.31 \pm 0.03$ & Keck/ESI (2017-01-21, \citetalias{senchynaUltravioletSpectraExtreme2017}) &  &  \\
        SB2 & $1.30 \pm 0.06$ & Keck/ESI (2017-01-20, \citetalias{senchynaUltravioletSpectraExtreme2017}) &  &  \\
        J1119+5130 & $2.64\pm0.26$ & Keck/ESI (2017-02-21, \citetalias{senchynaExtremelyMetalpoorGalaxies2019}) &  &  \\
        SBSG1129+576 & $<0.42$ & Keck/ESI (2017-02-21, \citetalias{senchynaExtremelyMetalpoorGalaxies2019}) &  &  \\
        SB191 & $<0.14$ & Keck/ESI (2016-03-29, \citetalias{senchynaUltravioletSpectraExtreme2017}) &  &  \\
        SB111 & $3.94 \pm 0.13$ & Keck/ESI (2017-01-20, \citetalias{senchynaUltravioletSpectraExtreme2017}) & $0.79\pm0.36$ & \citet{izotovDetectionNeEmission2012} \\
        HS1442+4250 & $3.58 \pm 0.06$ & Keck/ESI (2017-02-21, \citetalias{senchynaExtremelyMetalpoorGalaxies2019}) &  &  \\
        J2251+1327 & $0.70 \pm 0.34$ & MMT/BC (2018-09-10) & $<0.60$ & MMT/BC (2018-09-10) \\
        SHOC595 & $<0.68$ & MMT/BC (2018-09-10) & $<0.62$ & MMT/BC (2018-09-10) \\

        \hline
    \end{tabular}

\end{table*}

\subsection{\chandra{} X-ray Imaging}
\label{sec:chandra}

All objects in our sample have publicly available \chandra{} ACIS-S imaging observations.
While detections have been published for several of these systems in the past, these prior studies generally do not include the coordinates of the identified X-ray sources and our sample also includes several galaxies covered by unpublished \chandra{} datasets.
This requires that we reanalyze the \chandra{} data in a uniform way.
We compare to these literature measurements where available in Appendix~\ref{appendix:casebycaseres}.
We first reprocessed the level 1 event files for these observations using the latest version of CIAO \citep[4.11, CALDB version 4.8.2][]{fruscioneCIAOChandraData2006}, focusing on the back-illuminated S3 chip which all targets were placed on.
We used \textsc{fluximage} to produce a clipped exposure map in the broad \chandra{} band, which we then used for source detection with a significance threshold of $10^{-6}$ run on the $\sqrt{2}$ series of pixel scales \citep[see e.g.][]{mineoXrayEmissionStarforming2012,brorbyXrayBinaryFormation2014}.

The spatial resolution afforded by \chandra{} is a key advantage in conducting a study connecting X-ray observations to optical data.
The astrometry of \chandra{} detections on the ACIS-S detector is accurate to $<1.4''$ in 99\% of cases in comparisons with ICRS optical counterparts\footnote{CIAO manual: \url{http://cxc.cfa.harvard.edu/cal/ASPECT/celmon/}.}.
Since we are concerned in this work with the potential impact of HMXBs on nebular line emission measured on spatial scales of $\sim 1''$ (Section \ref{sec:optspec}), we are not interested in X-ray sources significantly offset from the optical galaxy.
This is in contrast to much of the previous work on these galaxies, where significantly offset sources were included to account for the fact that HMXBs may be kicked up to hundreds of parsecs from their birthplace \citep[e.g.][]{kaaretDisplacementXraySources2004,zuoDisplacementXrayBinaries2010,brorbyXrayBinaryFormation2014}.
It is important to note that even with this resolution, we cannot unambiguously associate X-ray sources with the gas in which the observed \heii{} is excited.
At the distances of our targets, $1''$ corresponds to comoving physical distances of 50 pc to 3 kpc, larger than the sizes of individual star clusters in the local universe \citep[e.g.][]{meurerStarburstsStarClusters1995}.

We adopt the 99\% confidence positional accuracy limit of $1.4''$ as a limiting radius to establish association between the X-ray sources and the observed optical line emission, and consider separations between 1.4$''$--5$''$ on an individual basis (see Section~\ref{subsec:casebycaseres} and Appendix~\ref{appendix:casebycaseres}).
To compute net count rates, fluxes, and 68\% uncertainties for these detected sources in the 0.5--8.0 keV band, we ran \textsc{srcflux} on the filtered level 2 event files assuming an absorbed power law model with photon index $\Gamma= 1.7$ and a neutral absorbing column with density $N_H$ set to the Galactic value towards each system in the NRAO dataset of \citet{dickeyGalaxy1990} accessed by \textsc{colden}.
This photon power law is a reasonable approximation of the shape of the assumed intrinsic MCD spectrum over the 0.3--8.0 keV band, and is commonly assumed in flux measurements for similar samples \citep[e.g.][]{dounaMetallicityDependenceHighmass2015}.
Adopting a steeper index of $\Gamma=2.0$ \citep[corresponding to a softer spectrum, as in][]{mineoXrayEmissionStarforming2012,fabbianoPopulationsXRaySources2006a} would decrease the inferred fluxes by $12^{+1}_{-2}$\%.
If no sources were detected within 1.4\arcsec{} from the SDSS ICRS coordinates on which the optical spectra were centered, we consider the target undetected in X-rays.
In this case, we compute a 95\% confidence upper limit to the count rate and flux using \textsc{srcflux}.
In both cases, we convert the observed fluxes to luminosities in the same band using the distance to each galaxy (Table~\ref{tab:obssumm}).
These measurements and upper limits are presented in Table~\ref{tab:chandra}.

One concern in computing the X-ray fluxes is the possibility that uncertainties in the absorbing column density of neutral gas towards each source might impact on the inferred X-ray luminosities.
We do not have a robust way to estimate this internal absorption for these galaxies without X-ray spectral information or Ly$\alpha$ constraints \citep[e.g.][]{thuanChandraObservationsThree2004}, so we follow the standard procedure for such data and correct only for Galactic $N_H$ as described above.
Fortunately, the galactic absorbing column density towards our sources is uniformly low ($N_H<10^{21} \unit{cm^{-2}}$).
Increasing the assumed value of $N_H$ by a factor of five in the above-described \textsc{srcflux} measurements increases the resulting \lxb{} values by a median factor of only 1.15, suggesting that the effect of accounting for additional absorption internal to the target galaxies would likely be small.

\begin{figure*}
    \includegraphics[width=1.0\textwidth]{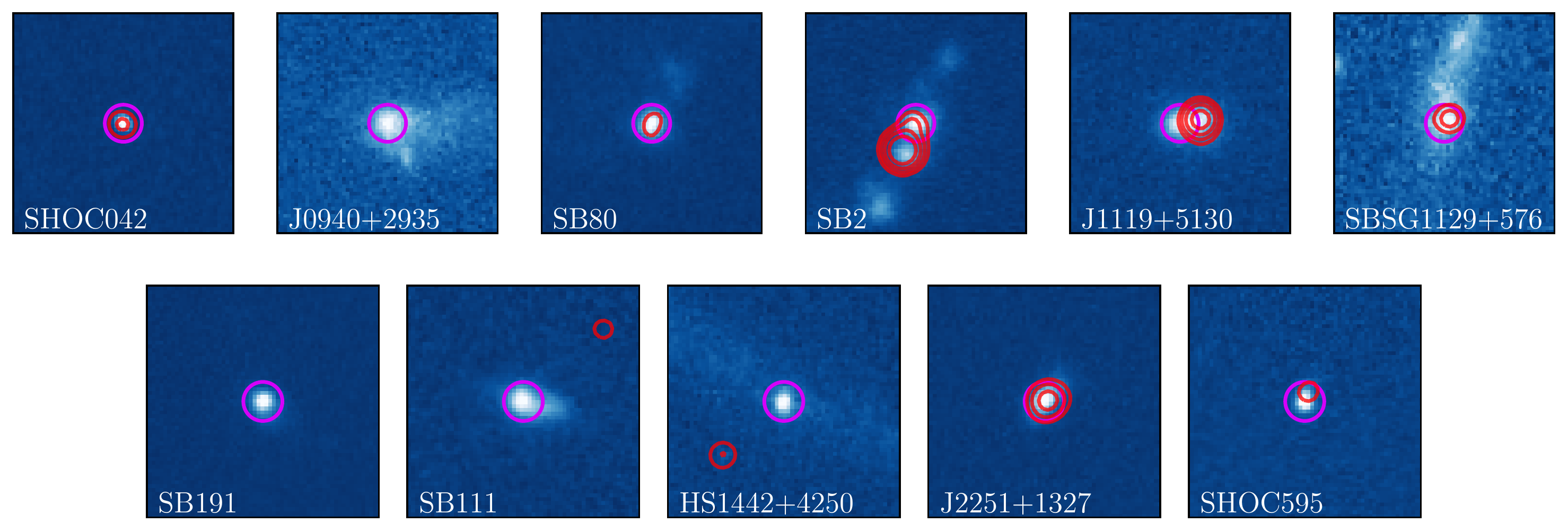}
    \caption{
        Optical and X-ray imaging for the eleven galaxies in our sample.
        The bluescale background image is a linear-scaled display of the SDSS $u$-band image probing light from recently-formed stars, and in red we overplot contour lines at count levels [0.25, 0.5, 1, 2] from the primary \chandra{} image after smoothing by a normalized $\sigma=1''$ Guassian.
        A purple circle with radius $2''$ centered on the optical spectral target (the brightest region in the $u$-band) is drawn in each case.
        \chandra{} X-ray sources are confidently detected within this optical radius in five cases, where it is possible for them to contribute to the observed nebular emission.
    }
    \label{fig:sdsschandra}
\end{figure*}

This procedure yields \chandra{} detections of seven of the galaxies in X-rays (Figure~\ref{fig:sdsschandra} and Table~\ref{tab:chandra}).
We discuss the results in more detail in Section~\ref{subsec:casebycaseres} and Appendix~\ref{appendix:casebycaseres}.

\begin{table*}
    \centering
    \caption{\label{tab:chandra} Summary of the \chandra{} ACIS-S observations analyzed in this work.
    We highlight the program ID and PI and the observation time for each.
    The separation between the nearest detected X-ray source and the optical spectroscopic target is given in both arcseconds and comoving kpc.
    We provide the galactic column density $N_H$ towards each source, and the count rate and derived luminosity if the target is detected with \chandra{} or the 95\% confidence upper limits if not (see Section~\ref{sec:chandra} and Appendix~\ref{appendix:casebycaseres}).
    }

\begin{tabular}{lccccccc}
\hline
Name & Chandra & Date & Exposure & $r_{\mathrm{sep}}$ & $N_H$ & Net count rate & \lxb{}\\ 
 & ObsID (PI) &  & ks & arcsec (comoving kpc) & $10^{20} \;\mathrm{cm^{-2}}$ & $\mathrm{s^{-1}}$ & $10^{38}\; \mathrm{erg \, s^{-1}}$\\ 
\hline
SHOC042& 16019 (Kaaret)& 2014-01-31& 22.7& 0.2 (0.72)& 2.67& $0.00089\pm0.00022$& $6.55 \pm 1.61 \times 10^3$\\ 
J0940+2935& 11301 (Prestwich)& 2010-01-16& 5.0& 60.3 (2.34)& 1.78& $<0.00054$ & $<0.34$\\ 
SB80& 19164 (Chandra)& 2018-01-03& 46.8& 0.2 (0.05)& 3.13& $0.00015\pm0.00007$& $4.94 \pm 2.23$\\ 
SB2& 19463 (Mezcua)& 2017-03-11& 14.9& 0.3 (0.03)& 4.25& $0.00078\pm0.00033$& $4.00 \pm 1.65$\\ 
J1119+5130& 11287 (Prestwich)& 2009-11-07& 11.7& 2.2 (0.24)& 1.16& $0.00847\pm0.00093$& $38.97 \pm 4.29$($<1.09$) \textdagger\\ 
SBSG1129+576& 11283 (Prestwich)& 2010-07-06& 14.8& 0.8 (0.09)& 1.00& $0.00177\pm0.00038$& $10.69 \pm 2.28$\\ 
SB191& 7092 (Swartz)& 2006-04-02& 2.0& 46.5 (2.25)& 2.46& $<0.00137$ & $<1.33$\\ 
SB111& 11290 (Prestwich)& 2010-07-26& 12.0& 11.1 (0.86)& 2.54& $<0.00023$ & $<0.59$\\ 
HS1442+4250& 11296 (Prestwich)& 2009-11-26& 5.2& 8.3 (0.44)& 1.54& $<0.00053$ & $<0.62$\\ 
J2251+1327& 13013 (Basu-Zych)& 2011-01-17& 19.6& 0.4 (0.53)& 4.75& $0.00216\pm0.00037$& $1.77 \pm 0.30 \times 10^3$\\ 
SHOC595& 17418 (Kaaret)& 2014-09-24& 13.6& 1.3 (3.59)& 4.72& $0.00076\pm0.00026$& $3.24 \pm 1.12 \times 10^3$\\ 
\hline
\end{tabular}

    \vspace{3mm}
    \begin{flushleft}
    \footnotesize{\textdagger The X-ray detection in this system is $>2''$ away from the optical spectroscopic target and likely does not contribute to the observed \heii{}, but is clearly associated with the galaxy; see Appendix~\ref{appendix:casebycaseres}. We consider this galaxy detected for the purposes of this paper, but note the 95\% confidence upper limit to the luminosity derived with \textsc{srcflux} centered on the spectroscopic target and note this caveat in Figure~\ref{fig:heiixr} as well.}
    \end{flushleft}

\end{table*}

\subsection{Results}
\label{subsec:casebycaseres}

With both optical and X-ray data in-hand, we can now build a picture of the X-ray properties of the galaxies in our sample.
For a detailed discussion of the results for each system, we refer the reader to Appendix~\ref{appendix:casebycaseres}.
Following the procedure described in Section~\ref{sec:chandra}, we detect X-ray emission cospatial with the star-forming clump targeted with optical spectroscopy in seven of the eleven galaxies in our sample (Table~\ref{tab:chandra} and Figure~\ref{fig:sdsschandra}).
The resulting measurements of \lxb{} range over three orders of magnitude, with the nearest sources revealing X-ray luminosities of 1--40$\times 10^{38}\; \mathrm{erg/s}$ and the more distant LBAs residing at 2--7$\times 10^{41} \; \mathrm{erg/s}$.
The available observations place strong upper limits on the X-ray luminosities of the undetected systems, constraining their luminosities to $\lesssim 10^{38}$ erg/s.
We find good agreement with the X-ray luminosities measured by \citet{brorbyXrayBinaryFormation2014,brorbyEnhancedXrayEmission2016} for the subset of our \chandra{} observations previously analyzed in these works, but our more stringent matching of the X-ray and optical coordinates reveals two key differences with earlier analysis.

\begin{figure}
    \includegraphics[width=0.45\textwidth]{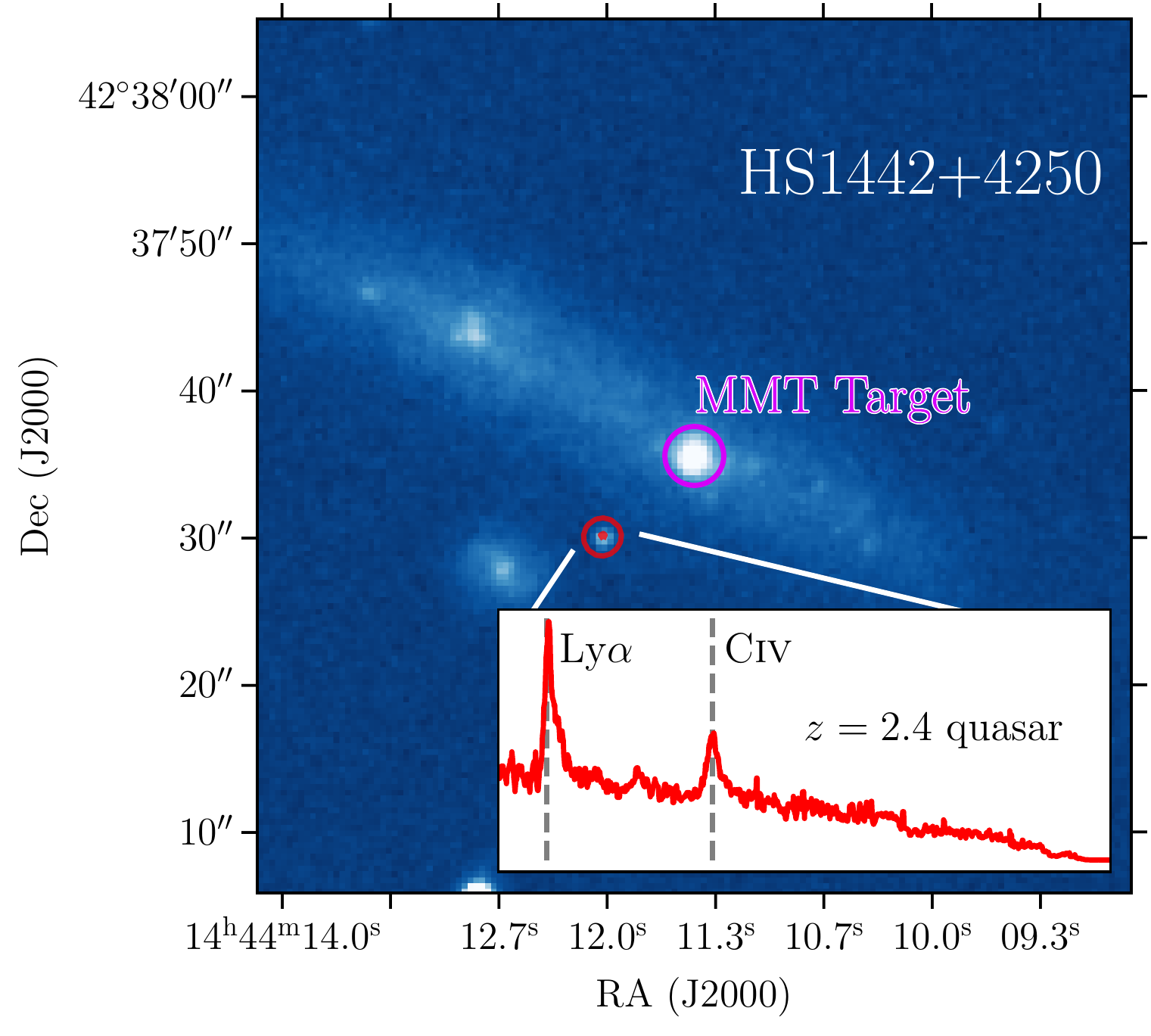}
    \caption{
        Follow-up spectroscopy of the X-ray source near HS1442+4250.
        The \chandra{} contours (red; same as Fig.~\ref{fig:sdsschandra}) correspond precisely to a faint point source visible in the SDSS $r$-band (background bluescale image) $8''$ away from the optical star-forming region in HS1442+4250 studied with previous deep spectroscopy (outlined with a purple $2''$-radius circle).
        A 10-minute MMT/BC spectrum (inset, red) reveals strong broad emission at 4130 and 5470 \AA{} (observed-frame), which we interpret as Ly$\alpha$ and \civ{} at $z=2.4$.
        In this case, the observed X-ray emission appears to be associated with a background AGN rather than an ejected HMXB associated with HS1442+4250.
    }
    \label{fig:hs1442xr}
\end{figure}

The undetected systems include two galaxies previously reported as HMXB hosts based upon the same \chandra{} data.
For both SB111 and HS1442+4250 \citep{brorbyXrayBinaryFormation2014}, we identified a nearby X-ray source, but found that it resided at a distance of 11\arcsec{} and 8\arcsec{} (respectively) from the center of the star-forming region targeted with ESI (Figure~\ref{fig:sdsschandra}).
This offset is far larger than the uncertainty associated with the \chandra{} coordinates ($<1.4\arcsec{}$ at 99\% confidence).
Even assuming the X-ray source resides at the same redshift, this places the object at a significant distance from the \hii{} region in which we observe nebular \heii{} (0.9 and 0.4 comoving kpc).
In the case of SB111, the nearest X-ray source has no SDSS optical counterpart.
However, the X-ray emission near HS1442+4250 is clearly associated with a faint $i=21$ point source in the SDSS image (Figure~\ref{fig:sdsschandra}).
As detailed in Appendix~\ref{appendix:casebycaseres}, we obtained an MMT/BC spectrum of this source and discovered broad emission lines consistent with \lya{} and \civ{} at $z=2.4$ (Figure~\ref{fig:hs1442xr}).
Thus, in this case the X-ray emission previously thought to belong to an ejected HMXB is actually associated with a background quasar.
In both cases, we conclude the X-ray emission cannot be physically associated with the observed nebular \heii{}.

Combined with the aperture SFR estimates from the Balmer lines (Section~\ref{subsec:basicprop}), these \chandra{} detections provide a measurement of the X-ray production efficiency of these galaxies.
These systems power luminosities \lxb{} ranging from $4\e{39}$ to $8\e{41}$ erg/s at a fixed SFR of 1 \unit{M_\odot/year} (Appendix~\ref{appendix:casebycaseres}).
Crucially, none of the galaxies exceeds $10^{42}$ erg/s per \unit{M_\odot/year}.
We discuss the implications of these measurements in the context of our photoionization model predictions in the following section.

\section{Discussion}
\label{sec:datajointres}

With the assembled set of eleven metal-poor galaxies with optical and X-ray constraints, we can test whether HMXBs are a plausible explanation for the observed nebular \heii{}.
In particular, we derived two key observational predictions from our photoionization modeling adopting the assumption that HMXBs provide the extra EUV ionizing radiation necessary to power \heii{} (Section~\ref{sec:model}).
First, we found that the HMXB spectrum could only boost the \heii{} luminosity relative to H$\beta$ to $\gtrsim 1\%$ at extraordinarily large X-ray luminosities per star formation rate of $>10^{42}$ erg/s at 1 \unit{M_\odot/year} (Figure~\ref{fig:heiinevnv_all}).
And second, our models indicate that this scenario should create a positive correlation between high X-ray luminosities at fixed SFR and strong high-ionization emission relative to H$\beta$ (especially high \heii{}/H$\beta$).

\begin{figure}
    \includegraphics[width=0.5\textwidth]{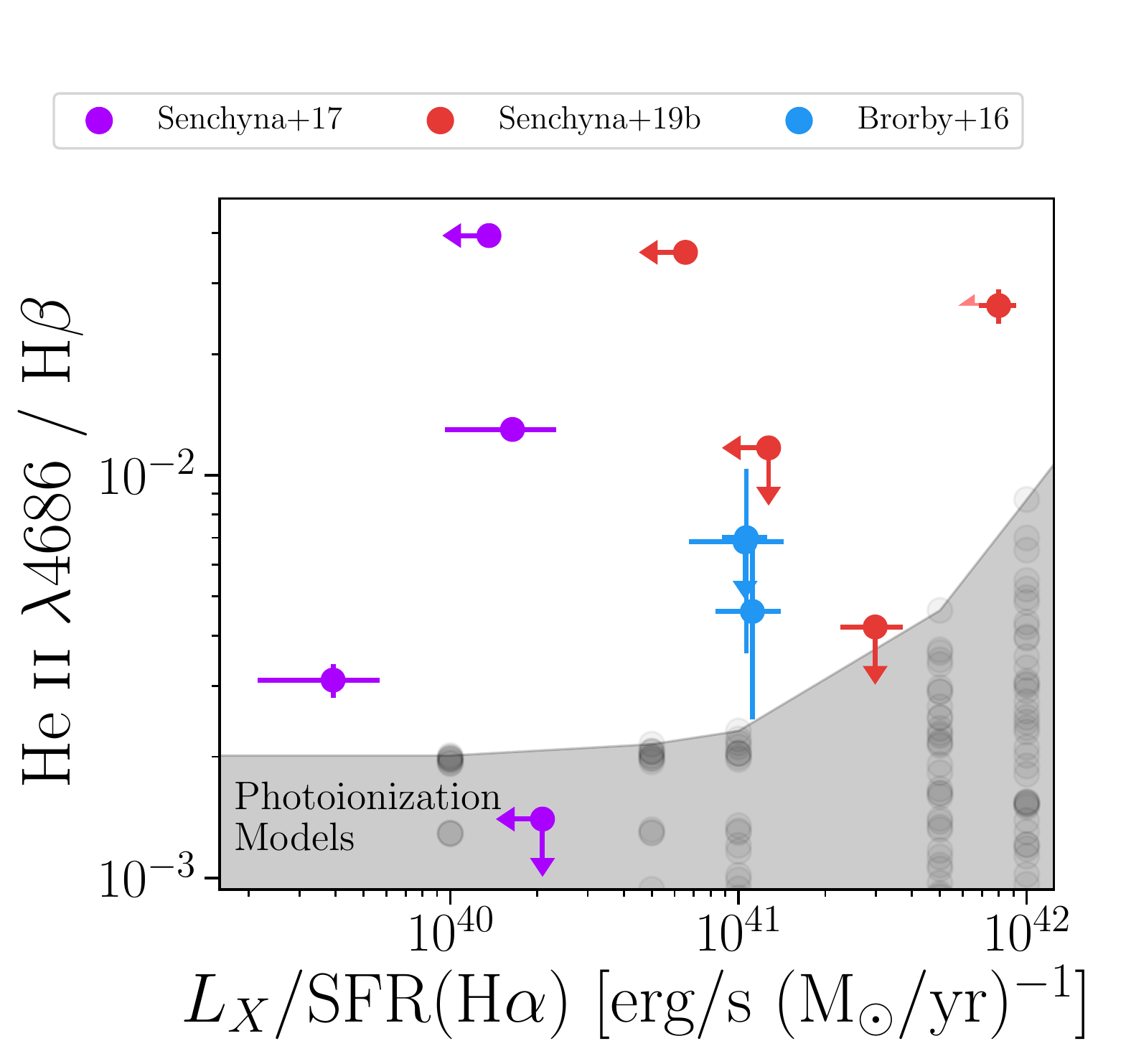}
    \caption{
        Observed and predicted \heii{}/H$\beta$ line ratios versus \lxb{}/SFR for our observational sample and our photoionization model grid.
        The observed galaxies detected in \heii{} lie uniformly in-excess of the maximal predictions of our photoionization model grid over the inferred range of X-ray production efficiencies.
        In addition, the most intense \heii{}-emitters show no indication of enhanced X-ray emission relative to galaxies with less extreme \heii{}.
        This comparison suggests that HMXBs are not the dominant source of \heii{} production in this sample of galaxies.
    }
    \label{fig:heiixr}
\end{figure}

First, we compare the X-ray production efficiencies attained by the galaxies in our sample to the results of our model grid.
In Figure~\ref{fig:heiixr}, we plot the flux ratio of \heii{} to H$\beta$ as a function of the X-ray luminosity per unit star formation rate measured for these eleven galaxies, alongside our full set of photoionization models.
Our targets with \heii{} detections are entirely disjoint from the photoionization modeling, powering \heii{} up to an order of magnitude stronger relative to H$\beta$ than expected.
For instance, consider SB2, in which we detect \heii{} at a high flux ratio with H$\beta$ of $0.0130\pm0.0004$.
This system is detected in X-rays as well, but at an X-ray production efficiency of only $1.47\pm 0.61 \e{40}$ erg/s for a star formation rate of $1\, \mathrm{M_\odot/year}$, two orders of magnitude lower than we estimate is required for the HMXB spectrum to reach the level of the stellar SED at the \heplus{}-ionizing edge.
Seven of our target systems are detected in \heii{}, including some of the highest values of \heii{}/H$\beta$ yet measured locally (up to 0.04), and yet all are found to be relatively inefficient producers of X-ray flux.
None reach an X-ray production efficiency of $10^{42}$ erg/s for a star formation rate of 1 $\mathrm{M_\odot/year}$, which we predict is the minimum required for HMXBs to begin powering strong \heii{}.
This indicates that HMXBs (with a standard MCD spectrum) in these galaxies do not produce sufficient flux at the $\mathrm{He^+}$-ionizing edge to account for the observed \heii{} emission.

Our photoionization models also make a clear prediction that if HMXBs are primarily responsible for $\mathrm{He^+}$-ionization, we should observe a correlation between \heii{}/H$\beta$ and the SFR-normalized X-ray luminosity, with the strongest \heii{} emission expected in galaxies most efficient in X-ray production.
Our observational results show no evidence for a relationship between these features.
In particular, the six galaxies with the strongest X-ray emission (exceeding $10^{41}$ erg/s in \lxb{} for a SFR of 1 $\mathrm{M_\odot/year}$) include only three \heii{} detections, with all but one revealing relatively-weak \heii{} (\heii{}/H$\beta \lesssim 0.01$).
The most prominent cases of \heii{} emission in our sample (\heii{}/H$\beta>0.025$, among the highest observed in any star-forming galaxies) occur in three systems: J1119+5130, SB111, and HS1442+4250.
The latter two of these are both confidently undetected in X-rays, with limiting X-ray production efficiencies as low as $<1.2\e{40}$ erg/s at $1\, \mathrm{M_\odot/year}$.
In the third case (J1119+5130), it is highly unlikely that the X-ray emission associated with the galaxy is capable of powering the observed \heii{} due to the sizable spatial offset from the spectroscopic aperture (Section~\ref{subsec:casebycaseres}).
Analogously to \heii{}, our models also suggest that the relative strength of \nev{} should correlate with higher X-ray production efficiencies if it is also powered by HMXBs.
While only three of our systems have \nev{} constraints, none of these three reveal detections of both X-ray emission and \nev{}.
The X-ray output of these systems appears entirely decoupled from the strength of high-ionization emission, contrary to our expectations assuming the two are physically related.

A potential source of concern in this analysis is HMXB variability, which could plausibly cause scatter relative to the models in Figure~\ref{fig:heiixr} if systems were observed with optical spectroscopy during an outburst and in X-rays during a period of quiescence.
Due to their nature as accretion-driven systems, HMXBs exhibit variability in both spectral shape and normalization during outbursts \citep[e.g.][]{mineoXrayEmissionStarforming2012,kaaretStateTransitionLuminous2013,brorbyTransitionXrayBinary2015}.
We see tentative possible evidence for variability in the X-rays in the multiple observations of SB80 and SHOC595 (Section~\ref{subsec:casebycaseres}).
However, the strongest nebular \heii{} emission in our sample (in SB111) was found at nearly identical strength relative to H$\beta$ in our 2017 Keck/ESI spectrum and in a 2011 MMT/BC spectrum \citep{izotovDetectionNeEmission2012,senchynaUltravioletSpectraExtreme2017}, suggesting that in contrast to the emission from HMXBs, the \heii{} emission in this object is not significantly variable.
In addition, in this scenario we would expect to see with equal likelihood observations of systems with enhanced \lxb{} due to being caught in an outburst and relatively low \heii{}/H$\beta$ (i.e.\ residing to the right of the models in Figure~\ref{fig:heiixr}), which we do not observe.
This argument applies also to the similar concern that geometrical beaming \citep[as we see some evidence for in ultraluminous X-ray sources, e.g.][]{kingUltraluminousXRaySources2001,kaaretUltraluminousXRaySources2017} may lead some systems to appear under-luminous in X-rays while still producing strong \heii{}; in this scenario we would also expect some systems to appear unexpectedly bright in X-rays relative to their \heii{} emission when observed down the beam, which we do not.
Even examining the larger sample of X-ray detected galaxies presented by \citetalias{brorbyEnhancedXrayEmission2016}, there are no instances of a measurement of an X-ray production efficiency in-excess of $10^{42}$ erg/s per $1 \, \mathrm{M_\odot/year}$ SFR in any local dwarf galaxy.
Thus, HMXB variability or beaming cannot readily explain the discrepancies we have discussed.

With HMXBs effectively ruled-out as the dominant source of \heii{} in these galaxies, we consider the possibility that the model stellar ionizing spectra may simply be underestimating the flux at the \heplus{}-ionizing edge.
Even with the latest generation of models for the expanding outer atmospheres and winds of massive stars incorporating line-blanketing and NLTE effects, the estimation of emergent flux beyond the \heplus{}-ionizing edge for a given star remains significantly uncertain \citep[e.g.][]{gablerUifiedNLTEModel1992,pulsAtmosphericNLTEmodelsSpectroscopic2005}.
In addition, there are large systematic uncertainties in predictions of the evolutionary tracks taken by stars at low metallicity, especially for those which undergo binary mass transfer.
In particular, stars stripped by binary mass transfer or structurally reshaped by high rotation rates can potentially provide substantial ionizing flux beyond 54.4 eV even at relatively low initial mass \citep[e.g.][]{szecsiLowmetallicityMassiveSingle2015,vinkWindsStrippedLowmass2017,gotbergIonizingSpectraStars2017,gotbergSpectralModelsBinary2018,kubatovaLowmetallicityMassiveSingle2019}.
Such stars have only recently been included in any form in population synthesis predictions \citep[][]{eldridgeEffectStellarEvolution2012,gotbergImpactStarsStripped2019a}, and may provide a natural explanation for both the strong metallicity dependence and lack of correlation with high specific star formation rates found for \heii{} in dwarf galaxies \citep[e.g.][]{senchynaPhotometricIdentificationMMT2019a,platConstraintsProductionEscape2019} as well as some of the peculiar \heii{} nebulae lacking coincident massive stars in the Local Group \citep[e.g.][]{pakullHighexitationNebulaeMagellanic2009,kehrigGeminiGMOSSpectroscopy2011}.
While some constraints can be placed on such stars from resolved work in the Local Group \citep[e.g.][]{smithExtremeIsolationWN32018}, extending calibrations of stellar models below the metallicity of the SMC in this way is challenging with current facilities given the limited number of appropriate resolved stellar populations \citep[e.g.][]{bouretNoBreakdownRadiatively2015,garciaOngoingStarFormation2019,evansFirstStellarSpectroscopy2019a}.
The study of nebular \heii{} in large samples of dwarf galaxies in the context of improved stellar population synthesis models may provide some of our only insight into and constraints on such stars, which are likely to exist but challenging to directly observe.

It is still possible that other non-stellar ionizing sources contribute to the $\mathrm{He^+}$-ionizing photon budget.
Fast radiative shocks driven by supernovae and stellar winds are another candidate origin for \heii{} and \nev{} emission \citep[e.g.][]{dopitaSpectralSignaturesFast1996,thuanHighIonizationEmissionMetaldeficient2005,izotovDetectionNeEmission2012,platConstraintsProductionEscape2019}.
While galaxies with strong \heii{} emission often lack clear signatures of significant shock-ionized gas \citep[e.g.][]{senchynaUltravioletSpectraExtreme2017,kehrigExtendedHeII2018,bergWindowEarliestStar2018}, a small shock contribution can boost \heii{} while keeping the other high-ionization UV lines consistent with observations in some cases \citep{platConstraintsProductionEscape2019}.
However, no predictive model of this shock contribution can yet account for the strong observed metallicity dependence in nebular \heii{} \citep[though a metallicity-dependent IMF or sSFR scaling could help explain this;][]{platConstraintsProductionEscape2019}.
In addition, soft and supersoft X-ray sources such as accreting white dwarfs identified in the Local Group can have effective temperatures $T_\mathrm{eff} \simeq 10^5$--$10^6$ K (corresponding to $\sim 40$--$400$ eV) and can potentially contribute to $\mathrm{He}^+$ ionization if they reside in a sufficiently dense ISM \citep[e.g.][]{remillardIonizationNebulaeSurrounding1995,distefanoLuminousSupersoftXRay2003,distefanoDiscoveryQuasisoftSupersoft2004,liuPuzzlingAccretionBlack2013a,woodsWhereAreAll2016,craccoSupersoftXRaySources2018a}.
Accreting intermediate-mass or supermassive black holes $>10^3 \unit{M_\odot}$ produce more flux at low energies than the $\leq 100$ \unit{M_\odot} black holes we consider here, and are potentially required to explain some of the most luminous soft X-ray sources discovered nearby \citep[e.g.][and references therein]{fabbianoPopulationsXRaySources2006a,kaaretUltraluminousXRaySources2017}.
Observations provide strong evidence that accreting massive black holes do reside in low-mass galaxies \citep[e.g.][]{reinesDwarfGalaxiesOptical2013,baldassareTilde50000Sun2015,sartoriSearchActiveBlack2015} and in some cases, even in galaxies classified as star-forming by optical emission line diagnostics \citep{reinesNewSampleWandering2020}.
Detailed study of many of these candidates in unresolved dwarf galaxies directly in the X-rays is challenging, however, due to their distance and relatively low luminosities \citep[though see e.g.][]{baldassareXrayUltravioletProperties2017a}.
Deeper investigation of fainter high ionization lines like \nev{} which probe photons at higher energies than 54.4 eV may provide new insight into these softer sources of EUV radiation, as well as stringently constraining their potential impact on \heii{}.

Regardless of their impact on \heii{}, X-ray binaries likely played a dominant role in shaping the thermal history of the very early universe.
Accurate calibration of models for HMXB populations at low metallicities will prove crucial for accurate interpretation of the 21-cm signal of reionization \citep[e.g.][]{furlanettoCosmologyLowFrequencies2006,pritchardConstrainingUnexploredPeriod2010, mesingerSignaturesXraysEarly2013,fragosEnergyFeedbackXRay2013a,fialkovObservableSignatureLate2014,mirochaDecodingXrayProperties2014,mirochaWhatDoesFirst2019}.
While much progress has been made in establishing the X-ray luminosity of low-metallicity stellar populations \citep{brorbyXrayBinaryFormation2014,dounaMetallicityDependenceHighmass2015,brorbyEnhancedXrayEmission2016,bluemEnhancedXrayEmission2019}, our results underscore the importance of expanding the sample of local galaxies with \chandra{} constraints.
Of the 8 XMPs with \chandra{} detections thus far \citep{brorbyXrayBinaryFormation2014}, we have shown that the X-ray source in one (HS1442+4250) actually corresponds to a background quasar, and another (SB111/[RC2] A1228+12) is offset enough to make association fairly unlikely.
Additional X-ray observations are needed to more confidently constrain the X-ray binary populations at the extremely low metallicities relevant for modeling the first galaxies.

\section{Summary}
\label{sec:summary}

Nebular spectroscopy promises powerful insight into the physical conditions and stellar populations of galaxies, and the origin of nebular \heii{} remains one of the most significant sources of tension in such analyses of nearby galaxies.
Since the first statement of this problem \citep{garnettHeIIEmission1991}, HMXBs have remained a popular proposed solution.
In order to assess the potential impact of HMXBs on \heii{} and high-ionization nebular line emission more generally, we constructed a photoionization model grid using a joint SED incorporating both stellar light and a model HMXB spectrum with variable normalization (Section~\ref{sec:model}).
To test the predictions of this grid observationally, we assembled a sample of eleven $z<0.2$ star-forming galaxies with both \chandra{} ACIS-S X-ray constraints and deep optical spectra capable of resolving nebular \heii{} (Section~\ref{sec:obs}).
Our main conclusions are summarized as-follows:
\begin{enumerate}
    \item Due to their typically high disk temperatures, HMXBs are relatively inefficient producers of $\mathrm{He}^+$-ionizing photons even compared to typical stellar populations with a strong break at the 54.4 eV ionizing edge (Figure~\ref{fig:sedcomp}).
    Our photoionization modeling reveals that the low-energy tail of a HMXB spectrum can contribute significantly to \heii{} production only for galaxies with very large X-ray luminosities for their star formation rate, exceeding $10^{42}$ erg/s for a $1\, \mathrm{M_\odot/year}$ SFR (Figure~\ref{fig:heiinevnv_all}).
    Our results also indicate that if HMXBs power \heii{}, we should find a positive correlation between \heii{}/H$\beta$ and the X-ray production efficiency \lxb{}/SFR among local galaxies.
    \item Significant care must be taken in establishing physical association between X-ray emission and dwarf galaxies.
    We find several instances in which X-ray emission which could be associated with ejected HMXBs is nevertheless too spatially-offset from the optical regions under study spectroscopically to be physically related to the measured high-ionization emission.
    In addition, we find one instance in which an offset X-ray source previously identified as an HMXB actually corresponds to a background object which we identify spectroscopically as a quasar at redshift $z=2.4$ (Figure~\ref{fig:hs1442xr}).
    \item Among our sample of nearby \heii{}-emitters, we find X-ray production is far less efficient than we predict is required for HMXBs to power \heii{}.
    In particular, our galaxies are uniformly less efficient at producing X-rays than the $10^{42}$ erg/s in \lxb{} normalized to $1\, \mathrm{M_\odot/year}$ lower limit that we find is required to power strong \heii{} (i).
    This indicates that the HMXBs present in these systems are too faint to produce significant power at the \heplus{}-ionizing edge relative to the stellar SED.
    \item We find no correlation between higher X-ray production efficiencies and stronger \heii{}/H$\beta$ in our sample.
    Of the three strongest \heii{}-emitters in our sample, only one is detected in X-rays, and even in this case the X-ray emission is likely too spatially offset to be related to the observed nebular emission.
\end{enumerate}

Our results suggest that HMXBs are not responsible for the nebular \heii{} emission detected in the metal-poor galaxies in our sample.
Continued observations and modeling of nebular emission from higher-ionization emission lines such as \nev{} will be necessary to further constrain the potential impact of other non-stellar sources such as shocks or softer accretion systems (Section~\ref{sec:datajointres}).
However, the possibility remains that \heii{} is indeed powered by very metal-poor stellar populations.
Both stellar evolutionary and atmosphere models below the metallicity of the SMC ($Z/Z_\odot<0.2$) remain entirely theoretical, and population synthesis predictions especially at the highest energies are subject to substantial systematic uncertainties.
Detailed spectroscopic observations of large samples of nearby metal-poor galaxies may provide one of our best opportunities to study the physics of low-metallicity stars.

\section*{Acknowledgments}

The authors would like to thank Kevin Hainline and Juna Kollmeier for helpful conversations and comments.
This research has made use of data obtained from the Chandra Data Archive and the Chandra Source Catalog, and software provided by the Chandra X-ray Center (CXC) in the application package CIAO.
Some of the observations reported here were obtained at the MMT Observatory, a joint facility of the University of Arizona and the Smithsonian Institution.
Some of the data presented herein were obtained at the W.M. Keck Observatory, which is operated as a scientific partnership among the California Institute of Technology, the University of California and the National Aeronautics and Space Administration.
The Observatory was made possible by the generous financial support of the W.M. Keck Foundation.
The authors wish to recognize and acknowledge the very significant cultural role and reverence that the summit of Mauna Kea has always had within the indigenous Hawaiian community.
We are most fortunate to have the opportunity to conduct observations from this mountain.

D.\ P.\ S.\ acknowledges support from the National Science Foundation through the grant AST-1410155.
J.\ M.\ acknowledges support from a CITA National Fellowship.

An allocation of computer time from the UA Research Computing High Performance Computing (HPC) at the University of Arizona is gratefully acknowledged.
This research made use of \textsc{Astropy}, a community-developed core \textsc{python} package for Astronomy \citep{astropycollaborationAstropyCommunityPython2013}; Matplotlib \citep{hunterMatplotlib2DGraphics2007}; \textsc{Numpy} and \textsc{SciPy} \citep{jonesSciPyOpenSource2001}; the SIMBAD database, operated at CDS, Strasbourg, France; and NASA's Astrophysics Data System.

\bibliographystyle{mnras}
\bibliography{zoterolib}

\appendix

\section{Galaxy-by-Galaxy Analysis}
\label{appendix:casebycaseres}

Correctly interpreting both high-ionization nebular emission lines and X-ray imaging constraints requires careful analysis on an individual galaxy basis.
Here, we outline and discuss the observational results obtained for each galaxy in our sample and our synthesis of the X-ray and optical data.
Where available, we compare to previously-published X-ray luminosities derived from the same datasets by \citet{brorbyXrayBinaryFormation2014,brorbyEnhancedXrayEmission2016}.

\textbf{SHOC 042} \citep[spectoscopically characterized in the SDSS \hii{} galaxies with Oxygen abundances Catalog, SHOC;][]{kniazevStrongEmissionLine2004} is a compact star-forming galaxy at $z=0.167$ classified as an LBA on the basis of high FUV luminosity and surface brightness comparable to LBGs at high redshift \citep{heckmanPropertiesUltravioletluminousGalaxies2005,hoopesDiversePropertiesMost2007,overzierHubbleSpaceTelescope2008}.
It is the most distant galaxy in our sample at 805 Mpc, hosts relatively high metallicity gas at $12+\log\mathrm{O/H}=8.21$, and harbors the largest star formation rate in our sample: $\mathrm{SFR}(\mathrm{H}\alpha)=$ 5.9 $M_\odot/\mathrm{yr}$ (Section~\ref{subsec:basicprop}).
Nebular \heii{} is detected in this system with MMT/BC (Figure~\ref{fig:mmtheii}), with \heii{}/H$\beta=0.0046\pm 0.0021$.
In the \chandra{} ACIS-S image of this source (ObsID: 16019, PI: Kaaret), we detect a source 0.2\arcsec{} (0.7 comoving kpc) away from the center of the optical spectroscopic target (within our adopted 99\% confidence spatial offset tolerance of 1.4\arcsec{}: Section~\ref{sec:chandra}), with a derived luminosity in the 0.5--8 keV band of $6.6 \pm 1.6 \e{41}$ erg/s (Table~\ref{tab:chandra}).
This flux measurement is in agreement (within $1.5\sigma$) with that of \citet{brorbyEnhancedXrayEmission2016} using the same dataset \footnote{Since \citealt{brorbyEnhancedXrayEmission2016} extract flux in a large predefined ellipse rather than using \textsc{srcflux}, small offsets such as this are expected.}.
This corresponds to an X-ray production efficiency of $1.1\pm 0.3 \e{41}$ erg/s per \unit{M_\odot/year} of star formation.

\textbf{J0940+2935} \citep[KUG 0937+298 in][]{brorbyXrayBinaryFormation2014} is a tadpole-shaped dwarf galaxy at $z=0.0024$ (8 Mpc), the most nearby galaxy in our sample \citepalias{senchynaExtremelyMetalpoorGalaxies2019}.
This system is two orders of magnitude closer and presents an SFR more than four orders of magnitude lower than SHOC 042 ($10^{-3.6}$ \unit{M_\odot/year}), and harbors gas with an oxygen abundance clearly in the XMP regime ($12+\log\mathrm{O/H}=7.63\pm 0.14$).
The Keck/ESI spectrum of this source reveals no nebular \heii{} (\heii{}/H$\beta<1.2\e{-2}$), and the 5 ks \chandra{} exposure (ObsID: 11301, PI: Prestwich) reveals no significant detections within an arcminute of the galaxy ($2.3$ comoving kpc), in agreement with \citet{brorbyXrayBinaryFormation2014}.
Our 95\% confidence upper limit places this galaxy at an X-ray production efficiency of $<1.3\e{41}$ erg/s per \unit{M_\odot/year}.

\textbf{SB 80} is a bright star-forming region embedded in UGC 5189 at 46 Mpc distance, and hosts moderately metal-poor gas at $12+\log\mathrm{O/H}=8.24\pm 0.06$ \citepalias{senchynaUltravioletSpectraExtreme2017}.
Our deep Keck/ESI spectrum reveals both a broad stellar and a clearly-differentiated nebular component to \heii{}, with the narrow component measured at \heii{}/H$\beta=0.0031\pm 0.0003$.
This system was observed in seven different \chandra{} pointings targeting the luminous supernova 2010jl \citep[positioned $1.7'$ away from the star-forming region under study;][]{stollSN2010jlUGC2011} from 2010--2018.
The longest and most recent of these observations (ObsID: 19164; 46.8 ks, taken 2018-01-03) reveals a source at $0.2''$ measured separation from the optical target, with $\lxb{}=4.9\pm2.2 \e{38}$ erg/s ($2.0_{-0.8}^{+0.9}\times 10^{-15}$ $\mathrm{erg/s/cm^2}$).
The exposure times and target positions on-chip for the other six observations vary significantly, with some indication of lower flux at the $\sim 2\sigma$ level in one observation but no statistically-significant differences in the other cases
\footnote{
Three other observations covering SB 80 with exposure times $\sim 40$ ks exist: 13781 (40.51 ks; PI: Chandra, 2011-10-17), 13782 (39.54 ks; PI: Chandra, 2012-06-10), and 15869 (39.54 ks; PI: Chandra, 2014-06-01).
A coincident source is detected in 13781, with lower flux by a factor of ten at $2\sigma$ significance ($3.2_{-2.9}^{+4.5}\times 10^{-16}$ $\mathrm{erg/s/cm^2}$); 13782 also shows a detection, with flux consistent with 19164 ($1.8_{-0.7}^{+0.8}\times 10^{-15}$ $\mathrm{erg/s/cm^2}$); and in 15869, no source is detected within 20$''$, but the $1-\sigma$ flux distribution measured in 19164 is not excluded by the 90\% confidence upper limit.
Likewise, no spatially-consistent source is detected in the remaining three observations with exposure times ranging from 10--20 ks,  (11122 PI: Chandra, 11237 PI: Pooley, 13199 PI: Chandra), but the 90\% confidence limits are all $\geq 2.3\times 10^{-15}$ $\mathrm{erg/s/cm^2}$.
}.
For this work, we proceed with the highest-confidence flux measurement obtained in ObsID: 19164, which is also the highest measured flux for this source.
At a star formation rate of $0.13$ \unit{M_\odot/year}, the measured X-ray luminosity (Table~\ref{tab:chandra}) corresponds to $3.9\pm 1.8\e{39}$ \unit{erg/s} in the 0.5--8 keV band per \unit{M_\odot/year} of star formation.

\textbf{SB 2} is a dwarf galaxy at a distance of 19 Mpc dominated by four star-forming clumps.
SDSS and Keck/ESI spectroscopy of the brightest of these clumps reveals prominent nebular emission in the \heii{} $\lambda 4686$ \AA{} line, with \heii{}/H$\beta=0.0130\pm0.0006$ \citepalias{senchynaUltravioletSpectraExtreme2017} and a star formation rate from the Balmer lines of $10^{-1.61}$ \unit{M_\odot/year}.
This system was observed with a 15 ks \chandra{} observation (ObsID: 19463, PI: Mezcua, not yet published), revealing an X-ray source coincident with the optical clump observed with ESI ($r_\mathrm{sep}=0.3''$) with a computed luminosity of $\lxb{}=4.0\pm1.7\e{38}$ erg/s.
The optical clump to the southeast of the spectroscopically-targeted region also has an associated X-ray detection, with an even higher luminosity $\lxb{}=5.3\pm0.5\e{39}$ erg/s; but since this second source is 3.2$''$ away, we conclude it cannot contribute to the observed nebular flux and ignore it in this analysis.
This X-ray luminosity corresponds to a production efficiency of $1.6\pm0.7 \e{40}$ \unit{erg/s} per unit \unit{M_\odot/year}.

\textbf{J1119+5130} \citep[$\mathrm{[RC2]}$ A1116+51 in][]{brorbyXrayBinaryFormation2014} is a compact XMP ($12+\log\mathrm{O/H}=7.51\pm0.07$) at 22 Mpc forming stars at $\mathrm{SFR(H\alpha)} = 10^{-2.3}$ \unit{M_\odot/year} \citepalias{senchynaExtremelyMetalpoorGalaxies2019}.
This galaxy shows strong nebular \heii{} in a Keck/ESI spectrum at \heii{}/H$\beta=0.026\pm0.003$ \citepalias{senchynaExtremelyMetalpoorGalaxies2019}.
We detect an X-ray source in the 11.7 ks \chandra{} exposure covering this galaxy (ObsID: 11287, PI: Prestwich), with a luminosity of $\lxb{}=3.9\pm0.4\e{39}$ erg/s in good agreement with that reported in \citet[][after correcting for the small difference in adopted distances; 22 versus 20.8 Mpc]{brorbyXrayBinaryFormation2014}.
However, we find this source to be offset by $2.2''$ from the center of the star-forming region we target, which is larger than the expected astrometric uncertainties of the ACIS-S data (Section~\ref{sec:chandra}).
This places the X-ray source outside of the $1''$ slit and extraction box used with Keck/ESI, and corresponds to 0.2 kpc at the distance of this galaxy.
However, it overlaps the other side of this compact galaxy in the optical (Figure~\ref{fig:sdsschandra}).
Thus, while it is unlikely that this X-ray source contributes to the nebular \heii{} observed in the Keck/ESI spectrum, it does appear to be clearly associated with the galaxy.
We consider this galaxy detected with \chandra{}, but note this caveat in Table~\ref{tab:chandra} and in Figure~\ref{fig:heiixr}.
This detection yields an X-ray production efficiency of $8.0\pm1.2\e{41}$ erg/s for a star formation rate of 1 \unit{M_\odot/year}, the highest of our sample.
If we instead adopt the 95\% confidence upper limit on the X-ray flux obtained by applying \textsc{srcflux} to the \chandra{} image at the precise coordinates of the ESI target, the X-ray production efficiency is constrained to $<2.2\e{40}$ erg/s per \unit{M_\odot/year}, over an order of magnitude lower.

\textbf{SBSG1129+576} is a tadpole-like XMP at 25 Mpc dominated by a star-forming clump targeted with SDSS and Keck/ESI spectroscopy \citepalias{senchynaExtremelyMetalpoorGalaxies2019}.
These spectra reveal a very low oxygen abundance $12+\log\mathrm{O/H}=7.47\pm0.06$ and an SFR of $10^{-2.5}$ \unit{M_\odot/year}, but a nondetection of nebular \heii{}, \heii{}/H$\beta<0.004$ \citepalias[low signal-to-noise positive flux is visible at \heii{} in the Keck/ESI spectrum, but with FWHM significantly larger than the strong nebular lines, so we conclude it is stellar in origin:][]{senchynaExtremelyMetalpoorGalaxies2019}.
An X-ray source is also detected in this galaxy with \chandra{} (ObsID: 11283, PI: Prestwich), at a luminosity of $\lxb{}=11\pm2 \e{38}$ erg/s and at $0.8''$ separation from the center of the clump observed with ESI (within our adopted offset tolerance).
This source was also identified by \citet{brorbyXrayBinaryFormation2014} at a luminosity in good agreement with our measurement.
This measurement yields an X-ray production efficiency of $3.0\pm0.7\e{41}$ erg/s per \unit{M_\odot/year}.

\textbf{SB191} \citepalias{senchynaUltravioletSpectraExtreme2017} is a compact star-forming \hii{}-region embedded in the large spiral galaxy NGC 4204 at 10 Mpc distance, and hosts gas at a moderate oxygen abundance of $12+\log\mathrm{O/H}=8.30\pm0.07$ with star formation rate $10^{-2.2}$ \unit{M_\odot/year}.
This system is undetected in nebular \heii{}, with a purely broad stellar component identified with Keck/ESI.
The host galaxy NGC 4204 was observed with \chandra{}/ACIS-S in a single 2 ks pointing (ObsID: 7092, PI: Swartz) targeting a ULX candidate in another part of the galaxy \citep{swartzCompleteSampleUltraluminous2011}.
The star-forming region we have targeted is more than an arcminute away from the ULX candidate studied in this previous analysis, but still near enough to the center of the S3 chip for confident source detection.
This short exposure reveals no X-ray target detections within $45''$ ($>2$ comoving kpc) of SB191, placing a 95\% upper limit on the X-ray luminosity of $\lxb{}<1.3\e{38}$ erg/s (corresponding to $<2.1\e{40}$ erg/s per \unit{M_\odot/year}).

\textbf{SB111} ([RC2] A1228+12 in \citealt{brorbyXrayBinaryFormation2014}) is a blue compact dwarf at 16 Mpc, with low-metallicity gas at $12+\log\mathrm{O/H}=7.81\pm0.08$ just above the XMP cutoff and a star formation rate from the Balmer lines of $10^{-2.4}$ \unit{M_\odot/year} \citepalias{senchynaUltravioletSpectraExtreme2017}.
This system has one of the strongest nebular \heii{} lines detected among local star-forming galaxies, with \heii{}/H$\beta=0.039\pm0.001$ measured with Keck/ESI.
In addition, a deep MMT/BC spectrum was obtained for this target by \citet{izotovDetectionNeEmission2012}.
This spectrum revealed both \heii{} (with \heii{}/H$\beta=0.038\pm0.002$ in very good agreement with our later Keck/ESI measurement) and a $2\sigma$ detection of \nev{} $\lambda 3426$, with \nev{}/H$\beta=0.79\pm0.36$ (Table~\ref{tab:optspec}).
SB 111 was also targeted with \chandra{} (ObsID: 11290, PI: Prestwich), and an associated X-ray detection with $\lxb{}=3.4\e{38}$ erg/s was reported by subsequent analyses \citep{prestwichUltraluminousXRaySources2013,brorbyXrayBinaryFormation2014,dounaMetallicityDependenceHighmass2015}.
However, as noted in \citetalias{senchynaUltravioletSpectraExtreme2017} and confirmed by the reanalysis presented in this work, this X-ray source is offset from the optical galaxy by $11.1''$ (or 0.9 kpc comoving kpc; Figure~\ref{fig:sdsschandra}).
While it is possible this is an ejected HMXB associated with star formation in this galaxy, this X-ray source cannot in any case contribute to excitation of the observed nebular \heii{}.
For the purposes of this work, then, SB111 is undetected in X-rays, with $\lxb{}<0.6\e{38}$ erg/s ($<1.4\e{40}$ erg/s per \unit{M_\odot/year}).

\textbf{HS1442+4250} is a somewhat more extended XMP ($12+\log\mathrm{O/H}=7.65\pm 0.04$) which hosts a compact star-forming region with $\mathrm{SFR(H\alpha)}=10^{-3.0}$ \unit{M_\odot/year} that we observed with MMT and Keck spectroscopy in \citetalias{senchynaExtremelyMetalpoorGalaxies2019}.
The Keck spectroscopy reveals \heii{} at a similar intensity to that found in SB111, with \heii{}/H$\beta=0.0358\pm0.0006$, among the highest observed.
This galaxy was also in the sample studied with \chandra{} in \citet{brorbyXrayBinaryFormation2014}, where an X-ray source was reported with luminosity $\lxb{}=2.2\e{38}$ erg/s (ObsID: 11296, PI: Prestwich).
Our \chandra{} reanalysis revealed that as in the case of SB111, this X-ray source was significantly offset from the optical galaxy, at a distance of $8.3''$ (corresponding to 0.4 comoving kpc) where contribution to the optical nebular line flux is impossible.
In this case, we noted that the observed X-ray source was coincident with a faint point source in the SDSS image (at 14h44m12.01s, +42d37m30.05s; $g=25.1$, $i=21.03$).
On the night of April 15 2018, we observed this target with MMT/BC for 10 minutes.
The extracted spectrum revealed strong broad asymmetric emission at 4130 and 5270 \AA{}, which we interpret as Ly$\alpha$ and \civ{} $\lambda 1548,1550$ at redshift $z=2.4$.
At this redshift, the observed flux corresponds to a rest-frame $\sim 2$--$27$ keV band luminosity of $1.3\e{45}$ erg/s, just above the knee of the observed AGN luminosity function at $z\sim 2$ \citep{airdEvolutionXrayLuminosity2015}.
Thus in this case, the X-ray emission previously identified as associated with this galaxy appears to originate entirely from a distant quasar in chance projection rather than an ejected HMXB.
We place an upper limit on X-ray flux from the optical target of $\lxb{}<0.6\e{38}$ erg/s (corresponding to a production efficiency of $<6.6\e{40}$ erg/s per \unit{M_\odot/year}).

\textbf{J2251+1327} is a relatively distant (279 Mpc) LBA described in \citet{brorbyEnhancedXrayEmission2016} with direct-$T_e$ gas-phase metallicity $12+\log\mathrm{O/H}=8.29\pm 0.06$ and $\mathrm{SFR(H\alpha)}=1.7 \unit{M_\odot/year}$.
Our MMT/BC spectrum yields a detection of nebular \heii{}, with modestly large \heii{}/H$\beta=0.0070\pm 0.0034$; but no \nev{} (\nev{}/H$\beta<0.62$).
We find a clear X-ray detection in the \chandra{} ACIS-S image of this galaxy (ObsID: 13013, PI: Basu-Zych) separated by 0.4$''$ from the center of the optical target (0.5 comoving kpc, but within our spatial offset tolerance of 1.4\arcsec{}), with luminosity $\lxb{}=1.77\pm0.30\e{41}$ erg/s consistent with the measurement of \citet{brorbyEnhancedXrayEmission2016}.
This results in an estimate of the X-ray production efficiency of $1.1\pm 0.2 \e{41}$ erg/s per \unit{M_\odot/year}.

\textbf{SHOC 595} is another LBA at similarly large distance (589 Mpc) hosting the highest-metallicity gas in our sample along with SB191 ($12+\log\mathrm{O/H}=8.30\pm0.14$).
The \heii{} profile measured with MMT/BC for this galaxy is broader than the other nebular lines (Section~\ref{sec:optspec}), and we fail to detect a clear narrow contribution at $2\sigma$ (\heii{}/H$\beta<0.68$) or \nev{} (\nev{}/H$\beta<0.62$).
We detect a luminous X-ray source in the 13.6 ks \chandra{} observation (ObsID: 17418, PI: Kaaret), with luminosity $3.2\pm 1.1\e{41}$ erg/s, 1$\sigma$ lower than the measurement presented by \citet[][]{brorbyEnhancedXrayEmission2016}.
The source is offset by 1.3$''$ from the optical center of the galaxy, which corresponds to a large physical distance of 3.6 comoving kpc but is within our adopted $r_{\mathrm{sep}}$ cutoff representing the 99\% confidence interval of the \chandra{} positional accuracy (Section~\ref{sec:chandra}).
This galaxy was observed in 2 separate instances, but the system was not detected in the other 9.1 ks integration (ObsID: 16020), with a 90\% confidence limiting flux just below that measured in 17418.
We adopt the former flux measurement, yielding an X-ray production efficiency of $1.1\pm0.4 \e{41}$ erg/s per \unit{M_\odot/year}, and discuss the impact of this possible variability in Section~\ref{sec:datajointres}.

\label{lastpage}

\end{document}